\documentclass[sigconf]{acmart}
\usepackage{multirow}
\usepackage{algorithm}
\usepackage{algorithmic}
\usepackage{subfigure}
\usepackage{multicol}

\AtBeginDocument{%
  }

\copyrightyear{2025}
\acmYear{2025}
\setcopyright{acmlicensed}
\acmConference[MM '25]{Proceedings of the 33rd ACM International Conference on Multimedia}{October 27--31, 2025}{Dublin, Ireland}
\acmBooktitle{Proceedings of the 33rd ACM International Conference on Multimedia (MM '25), October 27--31, 2025, Dublin, Ireland}
\acmDOI{10.1145/3746027.3755014}
\acmISBN{979-8-4007-2035-2/2025/10}

\begin{document}

\title{ISDrama: Immersive Spatial Drama Generation through Multimodal Prompting}

\author{Yu Zhang}
\authornote{Equal contribution.}
\affiliation{%
  \institution{Zhejiang University, Shanghai AI Lab}
  \city{Hangzhou}
      \state{Zhejiang}
    \country{China}
}
\email{yuzhang34@zju.edu.cn}

\author{Wenxiang Guo}
\authornotemark[1]
\affiliation{%
  \institution{Zhejiang University}
  \city{Hangzhou}
      \state{Zhejiang}
    \country{China}
}
\email{guowx314@zju.edu.cn}

\author{Changhao Pan}
\authornotemark[1]
\affiliation{%
  \institution{Zhejiang University}
  \city{Hangzhou}
      \state{Zhejiang}
    \country{China}
}
\email{panch@zju.edu.cn}

\author{Zhiyuan Zhu}
\authornotemark[1]
\affiliation{%
  \institution{Zhejiang University}
  \city{Hangzhou}
      \state{Zhejiang}
    \country{China}
}
\email{schmittzhu@zju.edu.cn}

\author{Tao Jin}
\affiliation{%
  \institution{Zhejiang University}
  \city{Hangzhou}
    \state{Zhejiang}
    \country{China}
}
\email{jint_zju@zju.edu.cn}

\author{Zhou Zhao}
\authornote{Corresponding Author.}
\affiliation{%
  \institution{Zhejiang University, Shanghai AI Lab}
  \city{Hangzhou}
      \state{Zhejiang}
    \country{China}
}
\email{zhaozhou@zju.edu.cn}

\begin{abstract}
Multimodal immersive spatial drama generation focuses on creating continuous multi-speaker binaural speech with dramatic prosody based on multimodal prompts, with potential applications in AR, VR, and others. 
This task requires simultaneous modeling of spatial information and dramatic prosody based on multimodal inputs, with high data collection costs.
To the best of our knowledge, our work is the first attempt to address these challenges. 
We construct MRSDrama, the first multimodal recorded spatial drama dataset, containing binaural drama audios, scripts, videos, geometric poses, and textual prompts.
Then, we propose ISDrama, the first immersive spatial drama generation model through multimodal prompting. 
ISDrama comprises these primary components:
1) Multimodal Pose Encoder, based on contrastive learning, considering the Doppler effect caused by moving speakers to extract unified pose information from multimodal prompts.
2) Immersive Drama Transformer, a flow-based mamba-transformer model that generates high-quality drama, incorporating Drama-MOE to select proper experts for enhanced prosody and pose control.
We also design a context-consistent classifier-free guidance strategy to coherently generate complete drama.
Experimental results show that ISDrama outperforms baseline models on objective and subjective metrics.
The demos are available at \url{https://aaronz345.github.io/ISDramaDemo}. 
We provide the dataset and the evaluation code at \url{https://huggingface.co/datasets/AaronZ345/MRSDrama} and \url{https://github.com/AaronZ345/ISDrama}.
\end{abstract}



\maketitle

\section{Introduction}

\begin{figure}[ht]
\centering
\includegraphics[width=1\linewidth]{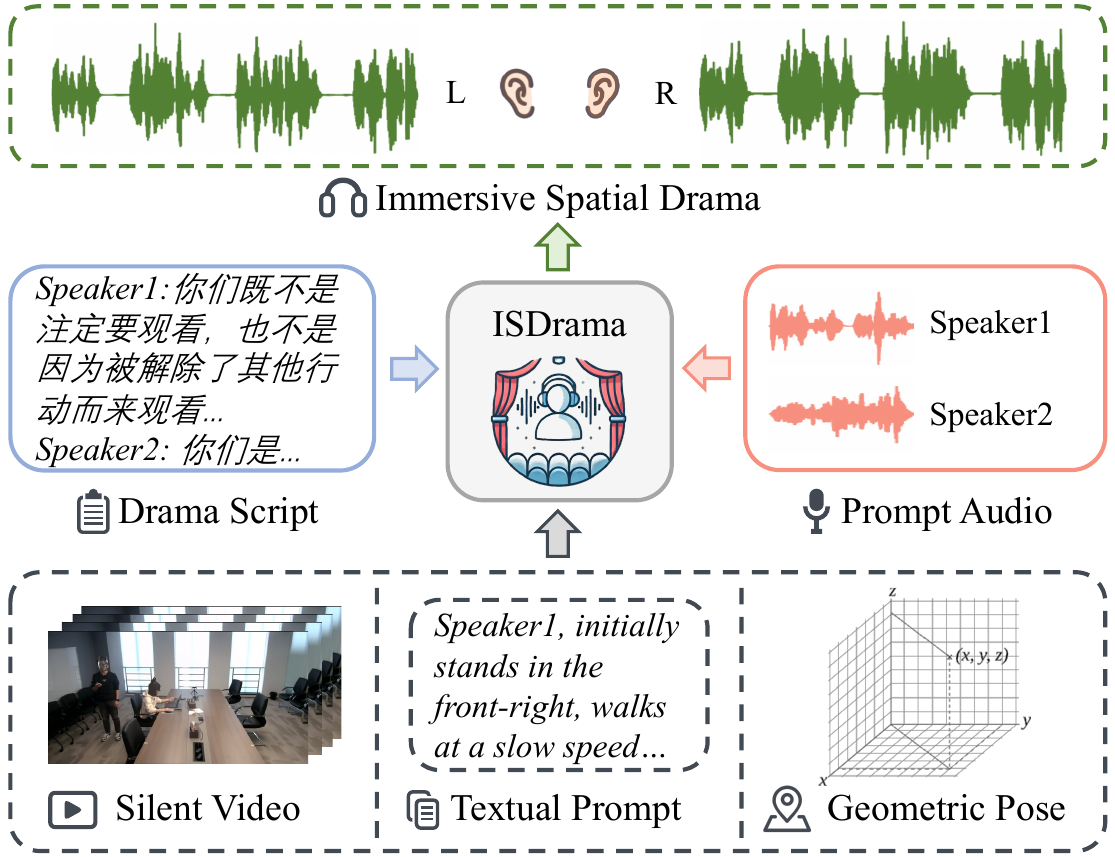}
\caption{ISDrama uses scripts as content, prompt audio to guide timbre, and spatial information from multimodal prompts, to create continuous multi-speaker binaural speech with dramatic prosody.}
\Description{ISDrama uses scripts as content, prompt audio to guide timbre, and spatial information from multimodal prompts, to create continuous multi-speaker binaural speech with dramatic prosody.}
\label{fig: intro}
\end{figure}

Binaural hearing provides localization cues through sound fields, enhancing the human spatial perception of the environment. 
This capability is critical for applications requiring deep immersion, like movies, VR, and AR \citep{fitria2023augmented,E240095,2020-3-397}. 
Compared to non-linguistic binaural audio \citep{sun2024both}, generating binaural speech is more challenging yet promising. 
Specifically, based on multimodal prompts from diverse contexts, generating continuous multi-speaker binaural speech with dramatic prosody can create immersive spatial drama.
This new task enhances storytelling, offering an immersive emotional and spatial engagement, and an integration of virtual and reality.

With deep learning advancements, progress has been made in synthesizing speech with prosody modeling \citep{ju2024naturalspeech,du2024cosyvoice} and generating binaural audio from monaural audio based on input multimodal prompts \citep{li2024cyclic,parida2022beyond}. 
Cascading these methods to generate continuous multi-speaker binaural speech with prosody modeling seems to be a viable approach.
However, this cascading method will disrupt the integration of prosody and spatial modeling, leading to mismatches and error accumulation.
This highlights the necessity for a unified framework for multimodal immersive spatial drama generation.

To date, one-stage multimodal immersive spatial drama generation remains an unexplored domain, focusing on creating continuous multi-speaker binaural speech with dramatic prosody.
As shown in Figure \ref{fig: intro}, this task requires inputting scripts as content and prompt audio to guide timbre, along with extracting \textbf{spatial information} in pose (including position, orientation, and movement speed) and scene from multimodal prompts (e.g., silent video, textual prompts, geometric poses) to cover a wider range of applications. 
Therefore, this task extends beyond video dubbing, enabling precise and flexible spatial control. 
Additionally, drama demands prosodic expressiveness that far exceeds normal speech.
Not only does \textbf{dramatic prosody} require learning accent and articulation from prompt audio, but it also integrates semantics to enhance the modeling of semantically aligned emotion and rhythm. 

Currently, multimodal immersive spatial drama generation encounters three major challenges:
1) \textbf{Lack of high-quality annotated recorded data.} 
Simulated data fails to capture the complex, dramatic prosody and the precise effects of real-world spatial scenes, positions, and orientations. 
While some datasets \citep{richard2021neural} use binaural devices to simulate human interaural phase difference (IPD) and interaural level difference (ILD), they suffer from limitations in scale, dramatic prosody, and multimodal prompts.
2) \textbf{Challenges in extracting unified pose representations from multimodal prompts.}
Silent video, geometric pose, and textual prompts provide spatial information, including positions, orientations, and movement speed for various scenarios. 
While some methods extract positional information from visual or positional inputs \citep{leng2022binauralgrad}, they cannot learn unified pose representations across more diverse scenarios.
3) \textbf{Difficulty in one-stage modeling dramatic prosody and spatial immersion.}
Existing monaural speech models \citep{du2024cosyvoice} struggle to simultaneously model dramatic prosody, which requires semantic alignment in the temporal-frequency domain, and spatial information, which spans the spatial-temporal dimensions.

To address these challenges, we first introduce \textbf{MRSDrama}, the first multimodal recorded spatial drama dataset, comprising binaural drama audios, scripts, videos, geometric poses, and textual prompts.  
The dataset includes 97.82 hours of speech data recorded by 21 speakers across three scenes.
Next, we propose \textbf{ISDrama}, the first immersive spatial drama generation model based on multimodal prompting.  
ISDrama generates high-quality, continuous, multi-speaker binaural speech with dramatic prosody and spatial immersion, driven by multimodal prompts.  
To extract a unified pose representation from multimodal prompts, we design the \textbf{Multimodal Pose Encoder}, a contrastive learning-based framework that encodes not only position and head orientation but also radial velocity, accounting for the Doppler effect caused by moving speakers.  
Meanwhile, we develop the \textbf{Immersive Drama Transformer}, a flow-based Mamba-Transformer model capable of generating immersive spatial drama effectively and stably.  
Within this model, we introduce Drama-MOE (Mixture of Experts), which selects the appropriate experts to enhance prosodic expressiveness and improve pose control.  
Then, we adopt a context-consistent classifier-free guidance (CFG) strategy to ensure the quality and coherence of complete drama generation.
We evaluate ISDrama on quality, speaker similarity, prosodic expressiveness, pose, angle, distance, IPD, and ILD.
Experimental results show that ISDrama outperforms baseline models on both objective and subjective metrics, demonstrating its ability to generate immersive spatial drama that adheres to physical principles while exhibiting rich prosodic variation.
Overall, our main contributions can be summarized as follows:

\begin{itemize}
\item We develop MRSDrama, the first multimodal recorded spatial drama dataset, accompanying videos, scripts, alignments, positions, and textual prompts.
\item We introduce ISDrama, the first immersive spatial drama generation model through multimodal prompting. We design the Multimodal Pose Encoder to extract pose information from multimodal inputs, while the Immersive Drama Transformer produces high-quality binaural speech.
\item Experimental results show that ISDrama outperforms baseline models on objective and subjective metrics.
\end{itemize}

\section{Related Work}

\noindent \textbf{Audio Spatialization with Multimodal Cues.}
In recent years, the development of deep learning has significantly advanced the exploration of spatial audio. 
Substantial progress has been made in sound source localization for binaural audio \citep{krause2023binaural,yang2022deepear,shimada2022multi,garcia2022binaural}. 
At the same time, the rise of multimodal research has spurred innovations in spatial audio synthesis.
Mono2Binaural \citep{gao20192} devises a deep convolutional neural network that leverages visual cues to convert monaural audio into binaural audio. 
Sep-Stereo \citep{zhou2020sep} enhances stereo audio by integrating audio-visual features. 
\citet{garg2021geometry} proposes a multi-task framework that learns geometry-aware features for generating binaural audio. 
CLUP \citep{li2024cyclic} jointly learns to localize visually sounding objects and generate binaural audio.
Beyond visual modalities, BinauralGrad \citep{leng2022binauralgrad} employs positional information for monaural-to-stereo audio generation, while scene depth maps have also been utilized \citep{parida2022beyond}. 
However, these methods rely on monaural audio input, limiting their applicability for complex and flexible generation tasks.
Recently, SpatialSonic \citep{sun2024both} employs a Diffusion Transformer to generate binaural audio from both text and image prompts. 
VISAGE \citep{kim2025visage} leverages CLIP visual features to generate first-order ambisonics directly from silent video frames.
However, these tasks focus on generating short audio clips lacking actual linguistic information.
ImmerseDiffusion \citep{heydari2025immersediffusion} proposes an end-to-end generative audio model conditioned on textual prompts of spatial, temporal, and environmental factors. 
However, it lacks support for multimodal prompts, binaural scenes, and expressive prosody modeling.
Generating continuous linguistic speech with unified scene and pose information extracted from more diverse prompt modalities remains an unresolved challenge.

\begin{table*}[ht]
\centering
\caption{Comparison of open-source recorded spatial speech datasets.
Multi-channel speech does not account for the intricate structures of human ears, whereas binaural speech incorporates natural IPD and ILD, ensuring an immersive hearing perception.}
\label{tab: data}
\begin{tabular}{lcccccc}
\toprule
\bf Dataset & \bf Hours & \bf Speakers & \bf Audio & \bf Paired Type \\
\midrule
Sweet-Home \citep{vacher2014sweet} & 47.3 & 71 & multi-channel & text, prompt \\
DIRHA \citep{ravanelli2015dirha} & 11 & 24 & multi-channel & text, pose \\
Voice-Home \citep{bertin2016french} & 2.5 & 3 & multi-channel &  text, pose, prompt \\
Binaural\citep{richard2021neural} & 2 & 8 & binaural & pose \\
\midrule
MRSDrama (Ours) & 97.82 & 21 & binaural & text, video, pose, prompt\\
\bottomrule
\end{tabular}
\end{table*}        

\begin{figure*}[ht]
\centering
\includegraphics[width=1\textwidth]{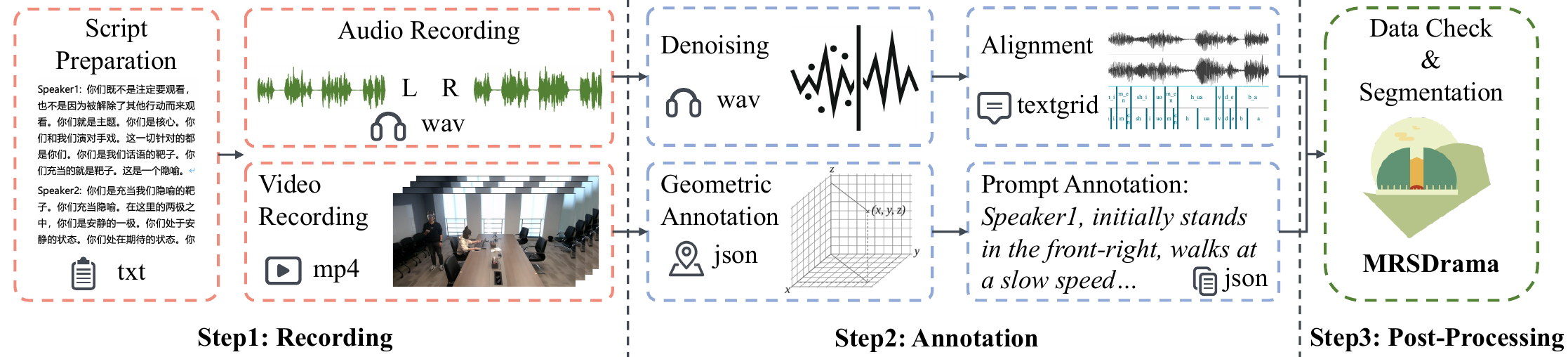}
\caption{The pipeline of MRSDrama data collection. 
Human double-checks exist in each process.
All data are desensitized.}
\Description{The pipeline of MRSDrama data collection.}
\label{fig: pipe}
\end{figure*}

\noindent \textbf{Speech Synthesis with Prosody Modeling.}
Prosody modeling aims to synthesize target speech with natural or emotional prosody, which is essential to generate expressive speech in a controlled manner and typically involves transferring prosody from prompt audio \citep{wagner2010experimental}.
\citet{skerry2018towards} is the first to integrate a prosody reference encoder into a Tacotron-based TTS system, enabling the transfer of prosody for similar-text speech.
Attentron \citep{choi2020attentron} introduces an attention mechanism to extract prosody from reference samples.
ZSM-SS \citep{kumar2021normalization} proposes a Transformer-based architecture with an external speaker encoder using wav2vec 2.0 \citep{baevski2020wav2vec}.
\citet{li2021towards} incorporates global utterance-level and local phoneme-level prosodic features in target speech. 
Daft-Exprt \citep{zaidi2021daft} employs a gradient reversal layer to enhance target speaker fidelity in prosody transfer.
Generspeech \citep{huang2022generspeech} incorporates the attention mechanism to capture multi-level prosody.
HierSpeech++ \citep{lee2023hierspeech++} generates F0 representation based on text representations and prosody prompts, while StyleTTS 2 \citep{li2024styletts} predicts pitch and energy based on a prosody predictor \citep{li2022styletts}.
NaturalSpeech 3 \citep{ju2024naturalspeech} employs factorized vector quantization to disentangle prosody.
CosyVoice \citep{du2024cosyvoice} incorporates x-vectors into an LLM to disentangle and model prosody.
FireRedTTS \citep{guo2024fireredtts} employs a semantic-aware speech tokenizer to encode speech style.
However, these models only focus on monaural speech synthesis and cannot simultaneously model dramatic prosody, which requires semantic alignment in the temporal-frequency domain, and spatial information, which spans the spatial-temporal dimensions.

\section{Dataset: MRSDrama}

Due to the high costs associated with multimodal spatial speech recording and annotation, as shown in Table \ref{tab: data}, existing open-source recorded datasets are inadequate for supporting the generation of multimodal immersive spatial drama. 
Meanwhile, simulated data cannot reflect the nuanced prosody or subtle acoustic effects of real-world spatial environments.
Therefore, we propose MRSDrama, the first multimodal recorded spatial drama dataset containing binaural drama audios, videos, scripts, geometric poses, and textual prompts.
Our dataset consists of 97.82 hours of speech data recorded by 21 speakers across 3 scenes. 
Figure \ref{fig: pipe} shows the construction pipeline.

\noindent \textbf{Recording.}
To construct ISDrama, we first source Chinese translations of renowned theatrical scripts through authorized channels, including \textit{Hamlet, Waiting for Godot, Thunderstorm, Hiroshima Mon Amour, Offending the Audience, etc.}
After a meticulous selection process, we enlist 21 expressive speakers, each provides consent to record for research purposes. 
Additionally, we select three scenes with varying sizes and acoustical effects to ensure diverse spatial effects.
During the recording process, speakers are instructed to read the script expressively while remaining in character and occasionally moving at a constant speed to create dynamic spatial effects. 
We use professional binaural recording equipment and sound cards to capture the audio, while synchronized video footage is recorded using cameras. 
All audio files are saved in WAV format with a 48 kHz sampling rate, and videos are captured at 24 fps.

\begin{figure*}[ht]
\centering
\includegraphics[width=1\textwidth]{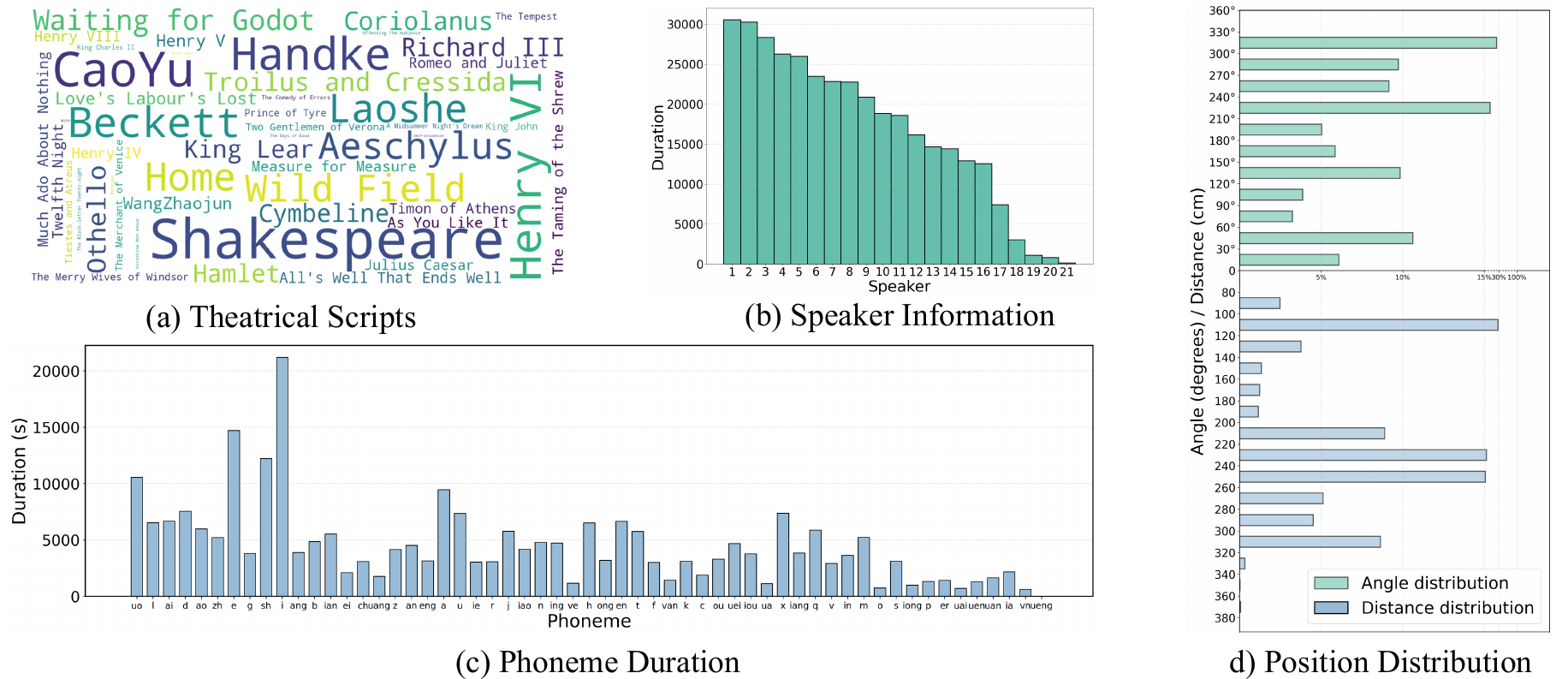}
\caption{The statistics of MRSDrama. 
The position distribution is plotted on the plane defined by listeners' direction and ears.}
\Description{The statistics of MRSDrama.}
\label{fig: stat}
\end{figure*}

\noindent \textbf{Annotation.}
We separately annotate the audio and video components.
1) For \textbf{Audio Annotation}, 
We first perform denoising using FRCRN \citep{zhao2022frcrn}. 
Subsequently, MFA \cite{mcauliffe2017montreal} is employed for a coarse phoneme-to-audio alignment between the original script and audio. 
Chinese phonemes are extracted by pypinyin. 
Next, annotators are asked to use Praat \cite{boersma2001praat} to refine the rough alignment, focusing on correcting word and phoneme boundaries and addressing erroneous words based on auditory perception. 
2) For \textbf{Video Annotation}, 
Speakers occasionally move between many fixed points in each scene.
Annotators are asked to record speakers' arrival times and point coordinates of each movement. 
They then measure these speakers' head orientation and mouth height while standing and sitting in each frame, extracting 3D position coordinates and quaternion orientation to form frame-level sound source poses.
Based on the annotated pose data, we then use GPT-4o \citep{achiam2023gpt} to generate textual prompts for each actor's line by combining the orientation, endpoint, direction, speed, and start time of each motion.
Meanwhile, annotators also label the camera pose (position and orientation) and scene prompts, including room sizes and acoustical effects.

\noindent \textbf{Post-Processing.}
To ensure data quality, we inspect the entire dataset, including the script, alignment, poses, and prompts.
Next, we segment the 97.82 hours of speech into 47,958 segments based on speaker transitions of each drama and a maximum duration setting of 16 seconds.
MRSDrama is the largest recorded spatial speech corpus to date and the first spatial drama dataset with multimodal annotations. 
It features continuous multi-speaker binaural speech with dramatic prosody accompanying rich modalities, making it suitable for various tasks like binaural localization and drama generation.
Figure \ref{fig: stat} presents the statistics of MRSDrama, showcasing the diversity across theatrical scripts, speakers, phonemes, and positions. 
This highlights the dataset's variety in content, timbre, and spatial information, demonstrating its potential for generalization.
For more details about MRSDrama, please refer to Appendix B.

\section{Method: ISDrama}

\begin{figure*}[ht]
\centering
\includegraphics[width=1\textwidth]{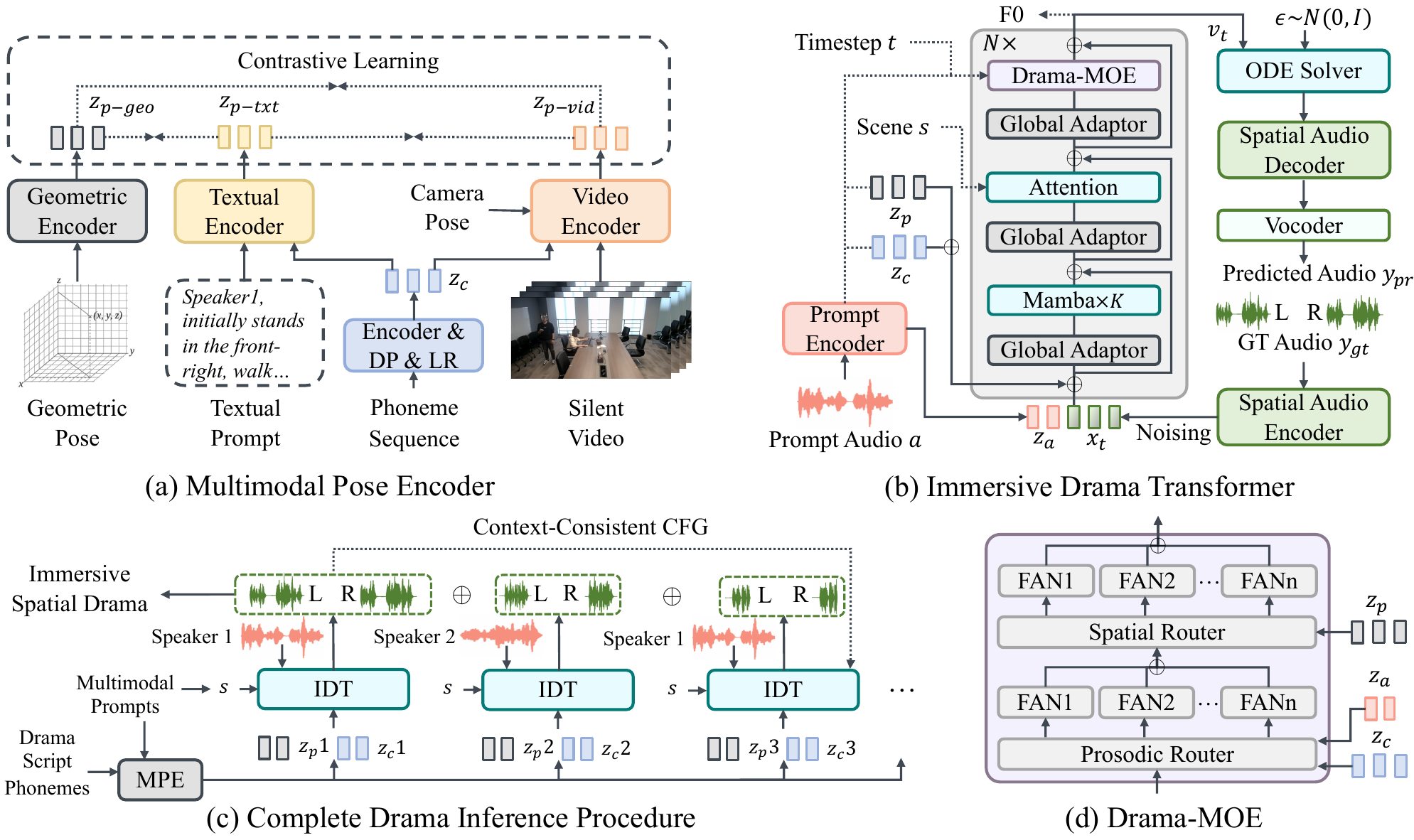}
\caption{The architecture of ISDrama.
In Figure (a), DP \& LR is duration predictor and length regulator.
In Figure (b), the ODE Solver, autoencoder decoder, and vocoder generate predicted audio from Gaussian noise with conditions during inference.
In Figure (c), MPE is the Multimodal Pose Encoder, while IDT is the Immersive Drama Transformer. 
MPE predicts content duration and segments inputs.
Then IDL coherently generates the complete drama.
In Figure (d), FAN is Fourier Analysis Networks.
}
\Description{The architecture of ISDrama.}
\label{fig: arch}
\end{figure*}

\subsection{Overview}

We aim to generate immersive spatial drama based on scripts, prompt audio for each speaker, and spatial information from multimodal prompts.
Let $y_{gt}$ represent one of the binaural speech in the ground truth drama, and $m_{gt} \in \mathbb{R}^{2 \times 80 \times T}$ represent the mel spectrogram, where $T$ denotes the target length.  
Typically, we segment target drama based on speaker transitions in the script.
As shown in Figure \ref{fig: arch}, since the autoencoder compresses $m_{gt}$ into $\hat{m_{gt}}$, the generation process is $\hat{m_{pr}} = G(\epsilon \mid C)$, where $\epsilon$ is Gaussian noise and $C$ represents the conditions. 
$C$ includes the corresponding content $c$, scene $s$, pose $p$ from multimodal prompts, and specified prompt audio $a$.
To encode unified pose embedding $z_p$ from multimodal prompts, we design the Multimodal Pose Encoder based on contrastive learning, accounting for the Doppler effect of moving speakers. 
In addition to the pose $p$ corresponding to $y_{gt}$, the content $c$ is segmented from the complete drama script. 
The pronunciation and semantics of $c$ are encoded as $z_c$.
Next, $z_p$ and $z_c$ are fed into the Immersive Drama Transformer, along with scene information $s$ (a video frame or textual description) and prompt audio $a$ for timbre, to generate the predicted binaural speech $y_{pr}$ through spatial and prosodic modeling.
Then, by combining different segments of $y_{pr}$, as shown in Figure \ref{fig: arch} (c), we can coherently generate the complete drama through a context-consistent CFG strategy.

\subsection{Multimodal Pose Encoder}
\label{sec: mpe}

To support a broader range of application scenarios, the generation of spatial drama must not only accommodate video dubbing but also offer flexibility in textual prompts and precise control over geometric poses, including positions, orientations, and movement speed. 
To address this, we design the Multimodal Pose Encoder, which predicts a unified pose embedding $z_p$ from multimodal prompts. 
As shown in Figure \ref{fig: arch}(a), our model encodes three types of multimodal prompts and embeds them into a unified space.

The correct phase estimation is crucial for binaural audio \citep{richard2022deep}. 
Therefore, for \textbf{geometric pose}, we not only encode the head orientation $ori$ and 3D coordinates of the sound source relative to left and right ears, $\vec{pos}_l$ and $\vec{pos}_r$ but also add the radial relative velocity according to Doppler effect \citep{gill1965doppler} for phase estimation \citep{liu2022dopplerbas}. 
Specifically, we calculate the 3D velocity vector $\vec{v}$ of the moving sound source in the Cartesian coordinate system, then decompose $\vec{v}$ into radial velocity components $v_{rad-l}$ and $v_{rad-r}$ in the spherical coordinate systems of left and right ears respectively by $\vec{v}_{rad} = \frac{\vec{pos} \cdot \vec{v}}{\| \vec{pos} \|} \cdot \hat{\mathbf{r}}$, where $\hat{\mathbf{r}} \in \mathcal{R}^1$ is the radial unit vector. 
Finally, we can encode and concatenate $(\vec{pos}_l, \vec{pos}_r, ori, v_{rad-l}, v_{rad-r})$ into $z_{p-geo}$.

To determine the target embedding length and considering the relationship between semantics, spoken speed, and motion, we encode the pronunciation and semantics of content $c$ and expand it with the predicted phoneme duration as $z_c$.  
Since the input script is organized in speaker transitions, we can compute predicted phoneme durations to segment the length for each target speech as shown in Figure \ref{fig: arch} (c). 
Then we employ a Transformer to predict $z_{p-txt}$ and $z_{p-vid}$, where $z_c$ serves as the input and leverages different modal conditions. 
For \textbf{textual prompts}, we use FLAN-T5 \citep{chung2024scaling} to encode text as the condition. 
For \textbf{silent video}, we encode and concatenate camera pose, mouth pixel sequences from Co-Tracker3 \citep{karaev2024cotracker3}, and embedding from CLIP \citep{radford2021learning} as the condition.

After obtaining pose embeddings of these three modalities, we design three types of contrasts for contrastive learning, each focusing on different aspects of the pose to explore diverse physical and spatial features. 
These include dynamic features (mobility, movement speeds, and movement direction), postural features (posture and orientation), and positional features (distance and angle).
We then use the contrastive objective \citep{radford2021learning} for training:
\begin{equation}
\begin{aligned}
\label{equ: contras}
& \mathcal{L}_{p1,p2} = -\frac{1}{2N}\sum_{i=1}^{N} (\log \frac{\exp ({sim}({z_{p1}}^i, {z_{p2}}^i)/\tau)}{\sum_{j=1}^{N} \exp ({sim}({z_{p1}}^i, {z_{p2}}^j)/\tau)} \\
& + \log \frac{\exp ({sim}({z_{p2}}^i, {z_{p1}}^i)/\tau)}{\sum_{j=1}^{N} \exp ({sim}({z_{p2}}^i, {z_{p1}}^j)/\tau)}),
\end{aligned}
\end{equation}
where $sim(\cdot)$ denotes cosine similarity.
The final total loss $\mathcal{L}_{contras}$ will be $\mathcal{L}_{p_{geo},p_{vid}} +\mathcal{L}_{p_{geo},p_{txt}} + \mathcal{L}_{p_{vid},p_{txt}}$.
After training, these three embeddings are aligned in the same space. 
The resulting unified $z_p \in \mathbb{R}^{2 \times H \times T} $, where $H$ denotes the hidden size, contains pose information, helping to model natural IPD and ILD of binaural speech.
For more details, please refer to Appendix C.4.

\subsection{Immersive Drama Transfomer}
\label{sec: idt}

\noindent  \textbf{Flow-based Mamba-Transformer.}
Spatial information and prosody in binaural speech are related to pose, scene, emotion, and rhythm. 
Their interaction necessitates the simultaneous modeling of both temporal-frequency and spatial-temporal, which becomes more complex with long sequences.
The flow matching method facilitates smooth transformations, promoting stable and rapid generation, while the Transformer architecture excels in sequence generation, making the flow-based Transformer particularly suitable for this task. 
Meanwhile, since Transformer tends to be computationally expensive for long sequences, we add Mamba blocks \citep{gu2023mamba} at an Attention-to-Mamba ratio of 1:K to balance memory usage, efficient training, and the ability to handle long sequences.

As shown in Figure \ref{fig: arch} (b), we add Gaussian noise $\epsilon$ to the autoencoder output $\hat{m_{gt}}$ to obtain $x_t$ at timestep $t$. 
We then add content embedding $z_c$ and pose embedding $z_p$ to $x_t$ and concatenate it with prompt audio embedding $z_a$. 
Therefore, using the self-attention mechanism and capability of Mamba blocks to capture long-range dependencies, Immersive Drama Transformer can effectively model content, pose, timbre, accent, and articulation.
Scene information $s$ (e.g., a video frame in silent video or a textual description for other inputs) is encoded and processed using the cross-attention mechanism to simulate the room's acoustic properties, such as the differences in acoustic effects caused by varying room sizes and acoustical effects.
Notably, we use the first block's output to predict binaural F0, providing extra supervision and additional input for subsequent blocks, helping to model dramatic prosody.
Furthermore, we employ RMSNorm \citep{zhang2019root} and design global adapters with AdaLN \citep{peebles2023scalable} to ensure training stability and global consistency in timbre and scene.
The final output vector field is trained as:
\begin{equation}
\label{equ: loss}
\begin{aligned}
&\mathcal{L}_{flow} = \mathbb{E}_{t, p_t(x_{t})} \left\| v_t(x_t, t | C; \theta) - (\hat{m_{gt}} - \epsilon) \right\|^2,
\end{aligned}
\end{equation}
where $p_t(x_{t})$ represents the distribution of $x_{t}$ at timestep $t$. 
For more details, please refer to Appendix A and C.6.

\noindent  \textbf{Drama-MOE.}
To enhance prosodic expressiveness and pose control, we propose Drama-MOE (Mixture of Experts), which selects suitable experts based on various input conditions. 
As shown in Figure \ref{fig: arch} (d), our Drama-MOE consists of two expert groups, each focusing on dramatic prosody and spatial information.
Prosodic MOE leverages prosody in aligned prompt audio embedding $z_a$ and semantics in content embedding $z_c$ to select suitable experts for fine-grained dramatic prosody modeling, such as an expert specialized in a happy, fast, high-pitched tone. 
$z_a$ is aligned with the inputs by a cross-attention model.
The Spatial MOE conditions on $z_a$, adjusting inputs to match the corresponding spatial information, such as sound source pose-induced changes and binaural differences in phase and loudness. 
It selects proper experts for different inputs, like one expert specialized in a sound source slowly approaching the listener from the far left-front direction.

Each expert in our design leverages Fourier Analysis Networks (FAN) \citep{dong2024fan} to decompose frequency components, enabling explicit modeling of binaural spatial dynamics while capturing speech periodicity, rhythm, and intonation.
The FAN layer is defined as:
\begin{equation}
    \phi(x) \triangleq [\cos(W_px)|| \sin(W_px)|| \sigma(B_{\bar{p}} + W_{\bar{p}}x)],
    \label{FANLayer}
\end{equation}
where $W_{p}, W_{\bar{p}} $, and $B_{\bar{p}} $ are learnable parameters.
FAN combines sinusoidal functions and a nonlinear activation, capturing periodic and non-periodic variations. 
This enhances the power for spatial-temporal and time-frequency auditory modeling, aiding in the synchronized modeling of speech prosody and spatial information.

Our routing strategies in Drama-MOE use dense-to-sparse Gumbel-Softmax \citep{nie2021evomoe}, enabling dynamic and efficient expert selection for each group.
Let $h$ be the hidden representation, and $g(h)_i$ denote the routing score for expert $i$.
To prevent overloading and ensure balance, we apply a load-balancing loss \citep{fedus2022switch}:
\begin{equation}
\label{equ: balance}
\begin{aligned}
&\mathcal{L}_{balance} = \alpha N \sum_{i=1}^{N} \left( \frac{1}{B} \sum_{h \in B} g(h)_i \right),
\end{aligned}
\end{equation}
where $B$ is the batch size, $N$ is the number of experts, and $\alpha$ a hyperparameter controlling regularization strength.
For more details and the algorithm, please refer to Appendix C.7.

\subsection{Complete Drama Inference Procedure}

For inference, users can input a complete script and specify each speaker's timbre with prompt audio $a$.  
We typically segment the target drama based on speaker transitions in the script. 
As shown in Figure \ref{fig: arch} (c), the Multimodal Pose Encoder predicts content duration and provides each target's embedding of pose $z_p$ and content $z_c$.  
Then, the Immersive Drama Transformer generates each target speech $y_{pr}$ from Gaussian noise $\epsilon$, conditioned on $z_c$, $z_p$, scene $s$, and $a$.  
To enhance generation quality and ensure contextual prosodic consistency, we design the context-consistent classifier-free guidance (CFG) strategy, which uses both prompt audio $a$ and the last predicted audio from the same speaker $y_{pr-last}$.
During inference, we modify the output vector field as:
\begin{equation}
\label{equ: cfg}
\begin{aligned}
& v_{cfg}(x, t|z_p;\theta) = \gamma \alpha v_t(x, t|a,C;\theta) +\\
& \gamma (1-\alpha) v_t(x, t|y_{pr-last},C;\theta)+ (1-\gamma) v_t(x, t|\varnothing,C;\theta),
\end{aligned}
\end{equation}
where $\gamma$ is the CFG scale that balances creativity and controllability, while $\alpha$ balances context consistency and controllability.   
Setting $\gamma = 3$ and $\alpha = 0.4$, we improve generation quality and add previously generated audio for the prosody consistency of the same speaker within a single drama act. 
This ensures coherence while preserving the timbre and accent of the original prompt audio.
Moreover, as prosody can be learned from previously generated audio in the same context, this approach also helps the modeling of semantically aligned prosody.
For more details, please refer to Appendix C.3. 

\section{Experiments}

\begin{table*}[ht]
\centering
\caption{Monaural speech quality comparison. 
For testing quality and speaker similarity, we evaluate single-sentence speech.}
\label{tab: mona}
\begin{tabular}{lcccccc}
\toprule
\multirow{2}{*}{\bf Method} & \multicolumn{3}{c}{\bf Objective Metrics} & \multicolumn{3}{c}{\bf Subjective Metrics} \\ \cmidrule(lr){2-4} \cmidrule(lr){5-7}
 & CER $\downarrow$ & SIM $\uparrow$ & FFE $\downarrow$ & MOS-Q $\uparrow$ & MOS-S $\uparrow$ & MOS-E  $\uparrow$ \\
\midrule
Ground Truth  & 2.58\% & - & - & 4.43 $\pm$ 0.12 & 4.41 $\pm$ 0.07 & 4.26 $\pm$ 0.11 \\
\midrule
Uniaudio & 4.21\% & 0.94 & 0.68 & 3.93 $\pm$ 0.12 & 4.06 $\pm$ 0.05 & 3.65 $\pm$ 0.18 \\
StyleTTS 2 & 4.19\% & 0.93 & 0.60 & 3.89 $\pm$ 0.04 & 4.02 $\pm$ 0.10 & 3.72 $\pm$ 0.09 \\
CoisyVoice & 3.95\% & 0.96 & 0.56 & 4.05 $\pm$ 0.06 & 4.19 $\pm$ 0.09 & 3.81 $\pm$ 0.21 \\
FireRedTTS & \bf 3.07 \% & 0.95 & 0.60 & 4.01 $\pm$ 0.17 & 4.11 $\pm$ 0.11 & 3.77 $\pm$ 0.10 \\
F5-TTS & 3.13\% & 0.96 & 0.55 & \bf 4.12 $\pm$ 0.16 & \bf 4.21 $\pm$ 0.08 & 3.86 $\pm$ 0.06 \\
\midrule  
ISDrama (ours) & 3.31\% & \bf 0.96 & \bf 0.34 & 4.06 $\pm$ 0.14 & 4.18 $\pm$ 0.11 & \bf 4.01 $\pm$ 0.09 \\
\bottomrule      
\end{tabular}
\end{table*}    

\begin{table*}[ht]
\centering
\caption{Binaural speech quality comparison.
We evaluate complete drama.
ANG and Dis denote angle and distance.
Spatialization refers to the generation of binaural audio directly from the GT monaural audio input based on the geometric pose.
}
\label{tab: bina}
\begin{tabular}{lcccccc}
\toprule
\multirow{2}{*}{\bf Method} & \multicolumn{4}{c}{\bf Objective Metrics} & \multicolumn{2}{c}{\bf Subjective Metrics} \\ \cmidrule(lr){2-5} \cmidrule(lr){6-7}
 & IPD MAE $\downarrow$ & ILD MAE $\downarrow$ & ANG Cos $\uparrow$ & DIS Cos $\uparrow$ & MOS-Q $\uparrow$ & MOS-P $\uparrow$ \\
\midrule
Spatialization & 0.007 & 0.043 & 0.58 & 0.79 & 4.09 $\pm$ 0.08 & 4.26 $\pm$ 0.16 \\
\midrule
Uniaudio & 0.012 & 0.060 & 0.38 & 0.64 & 3.65 $\pm$ 0.16 & 3.89 $\pm$ 0.11 \\
StyleTTS2 & 0.011 & 0.064 & 0.33 & 0.61 & 3.63 $\pm$ 0.05 & 3.81 $\pm$ 0.16 \\
CoisyVoice & 0.011 & 0.055 & 0.44 & 0.68 & 3.72 $\pm$ 0.13 & 4.02 $\pm$ 0.09 \\
FireRedTTS  & 0.010 & 0.051 & 0.42 & 0.65 & 3.78 $\pm$ 0.19 & 3.97 $\pm$ 0.17 \\
F5-TTS & 0.009 & 0.053 & 0.45 & 0.70 & 3.66 $\pm$ 0.14 & 4.11 $\pm$ 0.13 \\
\midrule  
ISDrama (geometric) & \bf 0.008 & \bf 0.046 & \bf 0.51 & \bf 0.75 & \bf 4.01 $\pm$ 0.14 & 4.18 $\pm$ 0.10 \\
ISDrama (video)  & 0.009 & 0.051 & 0.48 & 0.73 & 3.97 $\pm$ 0.11 & 4.09 $\pm$ 0.08\\
ISDrama (textual) & 0.011 & 0.055 & 0.43 & 0.68 & 3.95 $\pm$ 0.13 & \bf 4.41 $\pm$ 0.06\\
\bottomrule      
\end{tabular}
\end{table*}  

\subsection{Experimental Setup}

\noindent \textbf{Implementation Details.}
Mel-spectrograms are derived from raw binaural waveforms with a 48 kHz sample rate, 1024 window size, 256 hop size, and 80 mel bins. 
We use four Mamba-Transformer blocks, each of which includes three Mamba blocks. 
The flow-matching timestep is 1000 for training and 25 for inference with the Euler ODE solver. 
For the training procedure of Immersive Drama Transformer, we use 8 NVIDIA RTX-4090 GPUs with a batch size of 12K frames per GPU for 100K steps. The Adam optimizer is applied with a learning rate of $5 \times 10^{-5}$, $\beta_1 = 0.9$, $\beta_2 = 0.999$, and 10K warm-up steps.
Please refer to Appendix C.1 for more details.

\noindent \textbf{Evaluation Metrics.}
We perform both subjective and objective evaluations on the generated samples. 
To ensure a fair comparison with existing monaural speech synthesis models, we adopt monaural evaluation metrics and then evaluate binaural metrics for cascade-generated binaural speech with BinauralGrad \citep{leng2022binauralgrad}.
1) For \textbf{objective evaluation}, we use Character Error Rate (CER) and Cosine Similarity (SIM) to assess the content accuracy and speaker similarity with prompt audio. 
F0 Frame Error (FFE) is used to evaluate the quality of prosody modeling. 
Since existing binaural metrics are scarce and not suitable for our one-stage binaural speech generation task, we have designed several new metrics.
We extract Interaural Phase Difference (IPD) and Interaural Level Difference (ILD) from the binaural mel-spectrograms and compute MAE with GT. 
We also compute cosine similarity of angle and distance embedding extracted from SPATIAL-AST \citep{zheng2024bat} with GT for spatial evaluation.
2) For \textbf{subjective evaluation}, we conduct Mean Opinion Score (MOS), which is rated from 1 to 5 and reported with 95\% confidence intervals.
MOS-Q evaluates the synthesized quality (like naturalness, spatial perception, and coherence), MOS-S assesses speaker similarity in timbre and accent, and MOS-E measures the expressiveness of semantically aligned prosody. 
For binaural metrics, we employ MOS-P to evaluate the pose consistency between the multimodal prompt and the generated audio.
In the ablation study, we conduct Comparative Mean Opinion Score (CMOS).
For more evaluation details, please refer to Appendix D.

\begin{table*}[ht]
\centering
\caption{Ablation studies on Immersive Drama Transformer.
MPE denotes Multimodal Pose Encoder.}
\label{tab: abla}
\begin{tabular}{lcccccccc}
\toprule
\multirow{2}{*}{\bf Method} & \multicolumn{4}{c}{\bf Objective Metrics} & \multicolumn{4}{c}{\bf Subjective Metrics} \\ \cmidrule(lr){2-5} \cmidrule(lr){6-9}
 & CER $\downarrow$ & FFE $\downarrow$ & ANG Cos $\uparrow$ & DIS Cos $\uparrow$ & CMOS-Q $\uparrow$ & CMOS-S $\uparrow$ & CMOS-E $\uparrow$ & CMOS-P $\uparrow$ \\
\midrule
ISDrama (Geometric) & 3.31\% & 0.34 & 0.51 & 0.75 & 0.00 & 0.00 & 0.00 & 0.00 \\
\midrule
Geometric w/o MPE     & 3.35\% & 0.37 & 0.49 & 0.73 & -0.11 & -0.03 & -0.10 & -0.16 \\
Video w/o MPE         & 3.68\% & 0.41 & 0.38 & 0.63 & -0.39 & -0.19 & -0.29 & -0.52 \\
Textual w/o MPE       & 3.72\% & 0.44 & 0.35 & 0.61 & -0.50 & -0.20 & -0.32 & -0.57 \\
\midrule
w/o Drama-MOE         & 4.01\% & 0.49 & 0.39 & 0.65 & -0.46 & -0.27 & -0.40 & -0.52 \\
w/o Prosodic-MOE      & 3.55\% & 0.47 & 0.47 & 0.71 & -0.28 & -0.21 & -0.36 & -0.46 \\
w/o Spatial-MOE       & 3.48\% & 0.39 & 0.41 & 0.66 & -0.27 & -0.12 & -0.21 & -0.28 \\
w/o FAN               & 3.79\% & 0.46 & 0.40 & 0.66 & -0.43 & -0.24 & -0.32 & -0.39 \\
\midrule
w/o F0 pred           & 3.65\% & 0.46 & 0.46 & 0.69 & -0.36 & -0.19 & -0.41 & -0.48 \\
w/o Mamba             & 3.84\% & 0.49 & 0.43 & 0.65 & -0.30 & -0.08 & -0.27 & -0.39 \\
w/o CFG               & 3.73\% & 0.48 & 0.45 & 0.68 & -0.38 & -0.22 & -0.32 & -0.38 \\
\bottomrule      
\end{tabular}
\end{table*} 

\subsection{Results}

\textbf{Comparison of baseline models for monaural speech.}
To ensure a fair evaluation of the synthesized quality, speaker similarity to the prompt audio, and prosodic expressiveness, we compute monaural speech metrics after averaging the binaural speech generated by ISDrama across channels. 
For the baseline models, we employ several strong popular open-source speech synthesis models, including: 1) Uniaudio \citep{yang2023uniaudio}, 2) StyleTTS 2 \citep{li2024styletts}, 3) CosyVoice \citep{du2024cosyvoice}, 4) FireRedTTS \citep{guo2024fireredtts}, and 5) F5-TTS \citep{chen2024f5}. 
We use their open-source codes on GitHub and train them using our MRSDrama dataset in monaural speech format.
We also provide a detailed introduction for each baseline model in Appendix E.1.
To achieve a more precise evaluation of quality and speaker similarity, we conduct the assessment using only single sentences for all models.
Table \ref{tab: mona} shows that ISDrama performs well across all monaural metrics. 
In synthesized quality (CER, MOS-Q) and speaker similarity (SIM, MOS-S), it is comparable to the performance of the best speech synthesis baseline model, while it surpasses all baseline models in prosodic expressiveness (FFE, MOS-E). 
These well-performing results can be attributed to the integration of semantics to model dramatic prosody with semantically aligned emotion and rhythm in Drama-MOE, as well as F0 supervision during training.

\noindent\textbf{Comparison of Baseline Models for Binaural Speech.}
Based on the subjective and objective spatial metrics we designed, we train a commonly used monaural-to-binaural spatialization model, BinauralGrad \citep{leng2022binauralgrad}, on our MRSDrama dataset to convert the generated monaural speech from baseline models, conditioned on geometric poses, into binaural speech for evaluating spatial performance.
To better test the coherence of spatial information and the consistency of prosody, we concatenate sentences from the same generated drama for evaluation.
As shown in Table \ref{tab: bina}, with the same geometric pose input, the binaural speech generated by ISDrama outperforms all two-stage models across all spatial metrics.
The high MOS-Q shows coherence in continuous multi-speaker speech generation, which can be attributed to the effectiveness of our context-consistent CFG in generating complete drama.
In more flexible scenarios, including video and textual prompt inputs, ISDrama continues to deliver excellent performance, thanks to the unified pose representation extracted by the Multimodal Pose Encoder.
Although the textual prompt leads to lower objective metrics, it achieves a strong MOS-P score, which can be attributed to the general nature of the text descriptions. 
Specifically, the same textual description can generate multiple pose sequences, resulting in spatial information not fully aligned with the ground truth.

\subsection{Ablation Study}

\noindent\textbf{Multimodal Pose Encoder.}
Tables \ref{tab: bina} and \ref{tab: abla} present the results of feeding pose embeddings of different prompts into the Immersive Drama Transformer, with or without contrastive learning. We observe that the absence of contrastive learning has minimal impact on geometric poses. This is because the geometric pose embeddings include 3D orientation, quaternion orientation, and radial velocity, which provide sufficient information for accurate pose modeling. However, for silent video and textual prompts, omitting contrastive learning significantly reduces spatial performance. This highlights the importance of contrastive learning, as it enhances the model’s ability to generate unified embeddings from multimodal prompts, which is crucial for supporting a wide range of diverse and flexible immersive spatial drama application scenarios.

\noindent\textbf{Drama-MOE.}
Table \ref{tab: abla} shows the results of experiments where we removed the full Drama-MOE, eliminated individual expert groups, and replaced the FAN module with a simple linear layer. The results reveal that removing the full Drama-MOE leads to a decline in performance across all metrics. When examining the individual expert groups, we find that Spatial-MOE significantly affects spatial performance, while Prosodic-MOE influences both prosody expressiveness and speaker similarity. These observations suggest that Drama-MOE plays a key role in enhancing the modeling of both prosody and pose by utilizing specialized experts tailored to different spatial conditions and semantically aligned prosody. Additionally, we find that the FAN module outperforms the simple linear layer in all aspects, reflecting the benefits brought by frequency decomposition in capturing more nuanced features.

\noindent\textbf{Immersive Drama Transformer.}
We evaluate the effects of removing the supervision of F0, replacing Mamba with a memory-equivalent Transformer, and eliminating the context-consistent CFG. The results are shown in Table \ref{tab: abla}. It can be observed that F0 prediction is crucial for accurate prosody modeling, as its absence results in a significant loss of prosodic quality. The Mamba block, being lightweight, allows stacking more layers than a traditional Transformer with the same memory budget, which leads to further improvements in quality. Additionally, removing the context-consistent CFG reduces the consistency and prosodic expressiveness of the generated outputs. These findings underscore the effectiveness of these design choices in enhancing quality and prosody modeling capabilities of the Immersive Drama Transformer.

\section{Conclusion}

In this paper, we introduce a novel task, Multimodal Immersive Spatial Drama Generation, focusing on creating continuous multi-speaker binaural speech with dramatic prosody based on multimodal prompts.
To support this task, we present MRSDrama, the first multimodal recorded spatial drama dataset, comprising binaural drama audios, scripts, videos, geometric poses, and textual prompts. 
Then, we propose ISDrama,  the first immersive spatial drama generation model based on multimodal prompting.
To extract a unified pose representation from multimodal prompts, we design the Multimodal Pose Encoder, a contrastive learning-based framework.  
To generate immersive spatial drama effectively and stably, we develop the Immersive Drama Transformer, a flow-based Mamba-Transformer model, incorporating Drama-MOE, which selects proper experts to enhance prosodic expressiveness and pose control.  
Then, we design a context-consistent CFG strategy to coherently generate complete drama.
Experimental results show that ISDrama achieves better performance than baseline models.

\section*{Acknowledgements}

This work was supported by the National Key R\&D Program of China (2022ZD0162000) and the National Natural Science Foundation of China under Grant No.U24A20326.

\bibliographystyle{ACM-Reference-Format}
\bibliography{custom}

\clearpage
\appendix

\section{Preliminaries}
\label{app: pre}

\subsection{Doppler Effect}
\label{app: doppler}

The Doppler effect \citep{gill1965doppler} refers to the change in the frequency of a wave as observed by an observer when the source of the wave is in motion relative to it. 
Initially applied in radar systems, this effect helps to analyze the features of moving objects, such as targets of interest \citep{chen2006microdoppler}. 
The Doppler effect can be expressed as:
\begin{equation}
\label{eq:doppler}
\begin{aligned}
&f_o = \left( \frac{c}{c \pm v_{rad}} \right) f_s,
\end{aligned}
\end{equation}
where $c$ is the wave propagation speed, $v_{rad}$ is the radial velocity of the moving sound source, $f_s$ is the source frequency, and $f_o$ is the frequency received by the observer.

\subsection{Rectified flow-matching}
\label{app: flow}

In generative modeling, the true data distribution is denoted as $q(x_1)$, which can be sampled but lacks an accessible density function. 
Consider a probability path $p_t(x_t)$, where $x_0 \sim p_0(x)$ represents a standard Gaussian, and $x_1 \sim p_1(x)$ approximates the real data distribution. 
The core of the flow-matching approach \citep{liu2022flow} is to model this path directly, which is governed by the following ordinary differential equation (ODE):
\begin{equation}
\begin{aligned}
&\mathrm{d}x = u(x, t) \mathrm{d}t, \quad t \in [0, 1],
\end{aligned}
\end{equation}
where $u(x, t)$ denotes the target vector field, and $t$ is the time index. 
If we have access to the vector field $u$, it is possible to recover realistic data by reversing the flow. 
To approximate $u$, we use a vector field estimator $v(\cdot)$, and the flow-matching objective function is:
\begin{equation}
\begin{aligned}
&\mathcal{L}_{\mathrm{FM}}(\theta) = \mathbb{E}_{t, p_t(x)} \left\| v(x, t; \theta) - u(x, t) \right\|^2,
\end{aligned}
\end{equation}
where $p_t(x)$ denotes the distribution of $x$ at time $t$. 
When conditioning on additional $C$, the objective becomes the conditional flow-matching formulation \citep{lipman2022flow}:
\begin{equation}
\begin{aligned}
&\mathcal{L}_{\mathrm{CFM}}(\theta) = \mathbb{E}_{t, p_1(x_1), p_t(x | x_1)} \left\| v(x, t | C; \theta) - u(x, t | x_1, C) \right\|^2.
\end{aligned}
\end{equation}
The key idea behind flow-matching is to construct a direct transformation path from Gaussian noise to real data.
This is achieved by linearly interpolating between Gaussian noise $x_0$ and the real data $x_1$ to generate samples at time $t$:
\begin{equation}
\begin{aligned}
&x_t = (1 - t) x_0 + t x_1.
\end{aligned}
\end{equation}
Thus, the conditional vector field becomes $u(x, t | x_1, C) = x_1 - x_0$, and the rectified flow-matching (RFM) loss used for optimization is:
\begin{equation}
\label{equ: flow}
\begin{aligned}
&\left\| v(x, t | C; \theta) - (x_1 - x_0) \right\|^2.
\end{aligned}
\end{equation}
If the vector field $u$ is estimated correctly, realistic data can be generated by passing Gaussian noise through an ODE solver at discrete time steps. 
One effective method to solve the reverse flow is the Euler ODE:
\begin{equation}
\begin{aligned}
&x_{t + \epsilon} = x + \epsilon v(x, t | C; \theta),
\label{eq:euler}
\end{aligned}
\end{equation}
where $\epsilon$ is the step size. 
Flow-matching models typically require hundreds to thousands of training steps. However, the efficient linear interpolation approach significantly reduces this to 25 steps or fewer during inference, greatly enhancing computational efficiency.
Moreover, the seamless transition from noise to data ensures both stability and high-quality output, which is essential for generating complex data free of artifacts while maintaining consistency across diverse input conditions.

\subsection{Mamba}
\label{mamba}

Models based on Structured State Space (SSM), such as S4 and Mamba \citep{gu2023mamba}, draw inspiration from continuous systems that map a 1D sequence $x(t) \in \mathbb{R}$ to another sequence $y(t) \in \mathbb{R}$ through a hidden state $h(t) \in \mathbb{R}^\mathtt{N}$. 
In this system, the matrix $\mathbf{A} \in \mathbb{R}^{\mathtt{N} \times \mathtt{N}}$ serves as the state evolution parameter, while $\mathbf{B} \in \mathbb{R}^{\mathtt{N} \times 1}$ and $\mathbf{C} \in \mathbb{R}^{1 \times \mathtt{N}}$ are used for projection. 
The continuous system is:
\begin{equation}
\begin{aligned}
    h'(t) &= \mathbf{A}h(t) + \mathbf{B}x(t), \\
    y(t) &= \mathbf{C}h(t).
\end{aligned}
\end{equation}
S4 and Mamba are discrete-time versions of this continuous system, incorporating a timescale parameter $\mathbf{\Delta}$ to convert the continuous parameters $\mathbf{A}$ and $\mathbf{B}$ into discrete counterparts $\mathbf{\overline{A}}$ and $\mathbf{\overline{B}}$. 
A typical approach to perform this transformation is through zero-order hold (ZOH), which is:
\begin{equation}
\begin{aligned}
\label{eq:zoh}
    \mathbf{\overline{A}} &= \exp{(\mathbf{\Delta}\mathbf{A})}, \\
    \mathbf{\overline{B}} &= (\mathbf{\Delta} \mathbf{A})^{-1}(\exp{(\mathbf{\Delta} \mathbf{A})} - \mathbf{I}) \cdot \mathbf{\Delta} \mathbf{B}.
\end{aligned}
\end{equation}
After discretizing $\mathbf{A}$ and $\mathbf{B}$, the discrete-time system can be written as:
\begin{equation}
\begin{aligned}
\label{eq:discrete_lti}
    h_t &= \mathbf{\overline{A}}h_{t-1} + \mathbf{\overline{B}}x_{t}, \\
    y_t &= \mathbf{C}h_t.
\end{aligned}
\end{equation}
Finally, these models obtain the output through a global convolution operation, represented as:
\begin{equation}
\begin{aligned}
\label{eq:conv}
    \mathbf{\overline{K}} &= (\mathbf{C}\mathbf{\overline{B}}, \mathbf{C}\mathbf{\overline{A}}\mathbf{\overline{B}}, \dots, \mathbf{C}\mathbf{\overline{A}}^{\mathtt{M}-1}\mathbf{\overline{B}}), \\
    \mathbf{y} &= \mathbf{x} * \mathbf{\overline{K}},
\end{aligned}
\end{equation}
where $\mathtt{M}$ denotes the length of the input sequence $\mathbf{x}$, and $\mathbf{\overline{K}} \in \mathbb{R}^{\mathtt{M}}$ represents the structured convolutional kernel.

\section{Dataset Details}
\label{app: data}

\begin{figure*}[ht]
\centering
\includegraphics[width=1\textwidth]{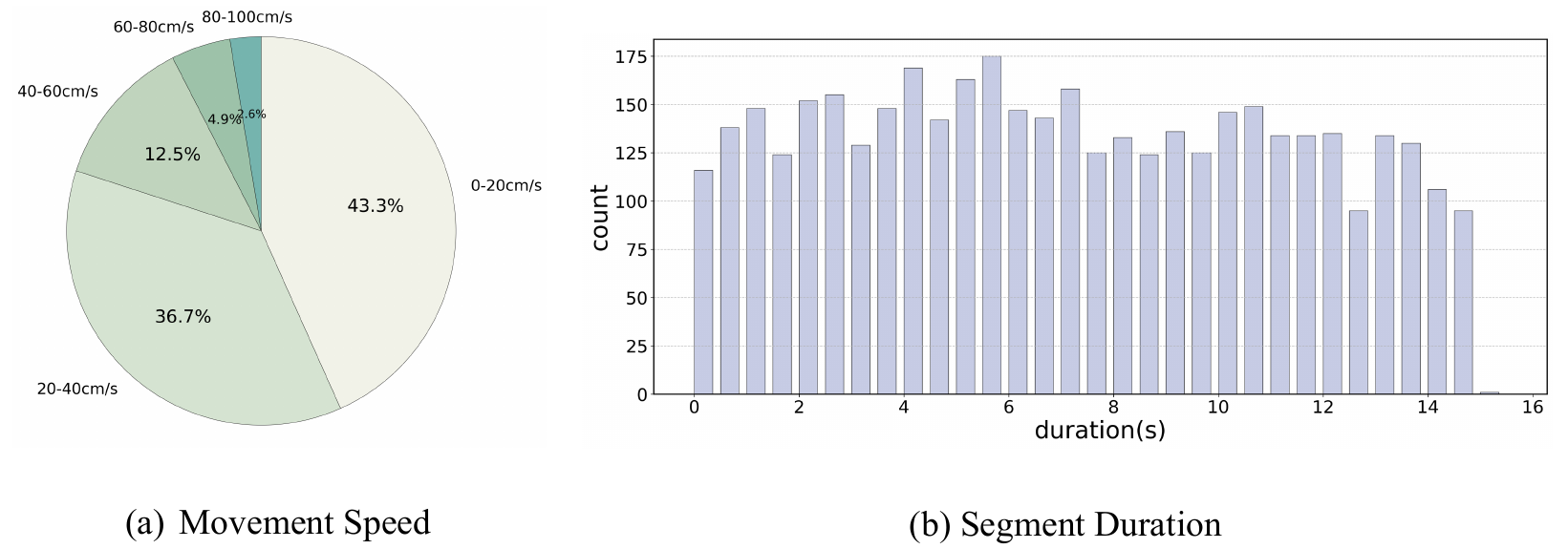}
\caption{The extended statistics of MRSDrama.}
\Description{The extended statistics of MRSDrama.}
\label{fig: stat2}
\end{figure*}

\begin{figure}[ht]
\vskip 0.2in
\centering
\includegraphics[width=1\linewidth]{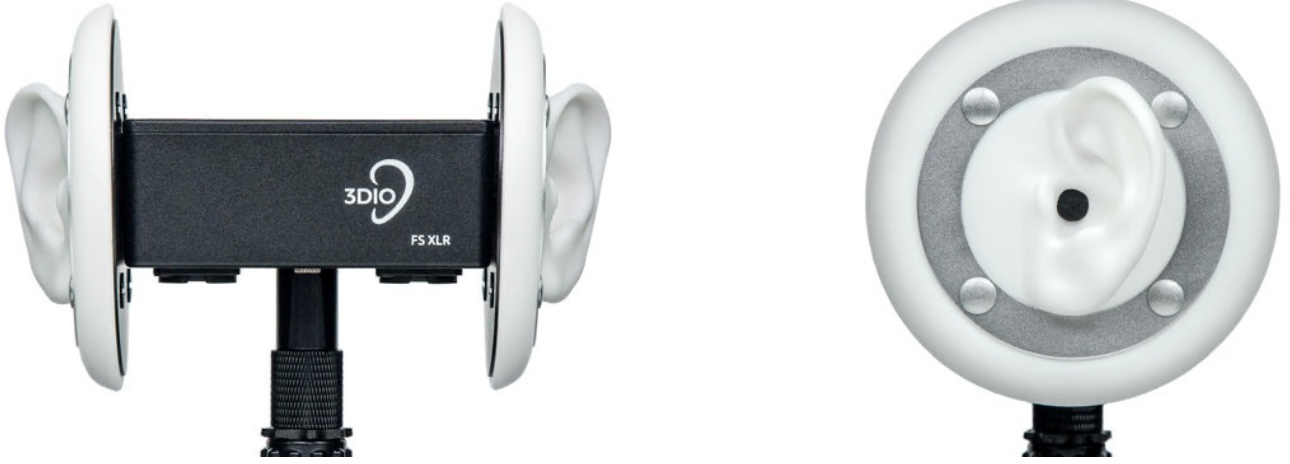}
\caption{The binaural recording device we used.}
\Description{The binaural recording device we used.}
\label{fig: device}
\end{figure}

Simulated data cannot accurately capture the intricate, dramatic prosody or the precise effects of real-world spatial scenes, positions, and orientations. Meanwhile, stimulated spatial audio datasets that expand on monaural audio data also have limited effectiveness in modeling the nuances of real-world scenes, positions, and orientations \citep{sarabia2023spatial,li2024sonicsim}.

As shown in Figure \ref{fig: device}, we utilize the 3Dio FS XLR Binaural Microphone\footnote{\label{foot: device}\url{https://3diosound.com/products/free-space-xlr-binaural-microphone}} connected to a Yamaha professional sound card to record binaural audio, effectively modeling interaural phase differences (IPD) and interaural level differences (ILD). The recordings are complemented by 24fps video captured using a camera. The scripts used in our study are sourced from authorized materials.
For the recording process, we design 5 to 12 fixed positions in each scene, covering various distances and angles. 
However, the routes and speeds are not predetermined, allowing speakers to move freely. Speakers may stand and walk through these points or sit on a chair at a designated position, which results in height variations.
This flexibility also aids in contrastive learning.
Speakers read from a script displayed on a screen that is not visible in the camera frame, with the requirement that the script has semantically aligned dramatic prosody. 
The recording of both audio and video for each dramatic act typically lasts between 3 and 15 minutes.
We hire the speakers at a rate of \$100 per hour, and they agree to make the video and audio data available for research purposes. 
We assure them that no facial information will be disclosed and then apply masks to their faces in the video using Adobe After Effects. 

Annotators label the 3D coordinates and quaternion orientations of the sound source (the speaker's mouth) based on the video. 
Using the annotated pose data, we employ GPT-4o \citep{achiam2023gpt} to generate textual prompts, such as: "Speaker 1 initially stands in the front-right, then walks at a slow pace to the far front-left..."
For binaural speech, we perform denoising using FRCRN \citep{zhao2022frcrn} and extract phonemes using PyPinyin.
Following previous Chinese audio annotation works \citep{zhang2024stylesinger, zhang2024tcsinger, zhang2024gtsinger, li2024robust, guo2025techsinger,zhang2022m4singer}. 
Coarse alignment between phonemes and audio is achieved using MFA \citep{mcauliffe2017montreal}. Annotators then refine the rough alignment using Praat \citep{boersma2001praat}, focusing on correcting word and phoneme boundaries and addressing erroneous words based on auditory perception. Each step is double-checked by human annotators to ensure accuracy.
We hire annotators at a rate of \$30 per hour, and they consent to their contributions being used for research purposes.
Finally, the dataset is segmented into 47,958 segments according to speaker transitions, with each segment having a maximum duration of 16 seconds.
Figure \ref{fig: stat2} shows more statistics of MRSDrama on movement speeds and segment duration.

\section{Model Design}
\label{app: mod}

\subsection{Model Configuration}
\label{app: config}

For the Multimodal Pose Encoder, we use a hidden size of 768. 
Our semantic information is modeled using BERT (base) \citep{devlin2018bert}, which preserves temporal length.
The duration decoder adopts the FastSpeech architecture \citep{ren2019fastspeech}. For the geometric encoder, we employ a Conv1D layer with a kernel size of 5. 
For the video and textual encoders, the Transformer model utilizes 8 transformer layers and 8 attention heads, with a hidden size of 768. 
The total number of parameters is 23.32M for each model. 
We obtain CLIP embeddings at 4 frames per second (FPS).
During training, we use the Adam optimizer with a learning rate of $1 \times 10^{-4}$, $\beta_1 = 0.9$, $\beta_2 = 0.999$, and 10K warm-up steps.

For the spatial audio encoder and decoder, we use a Variational Autoencoder (VAE) architecture \citep{kingma2013auto}. 
Mel-spectrograms are derived from binaural waveforms with a 48 kHz sample rate, 1024 window size, 256 hop size, and 80 mel bins. 
We use HiFi-GAN \citep{kong2020hifi} as the vocoder to synthesize waveforms from mel-spectrograms.
Specifically, the model consists of 3 layers for both the encoder and decoder, with a hidden size of 384 and a Conv1D kernel size of 5. The binaural mel-spectrogram with dimensions $B, 2, 80, T$ is compressed into dimensions $B, 40, T/4$, facilitating further processing by the Transformer. During training, batches of fixed length are used, consisting of 3000 mel-spectrogram frames. The Adam optimizer is employed with a learning rate of $1 \times 10^{-4}$, $\beta_1 = 0.9$, $\beta_2 = 0.999$, and 10K warm-up steps.

For the Immersive Drama Transformer, we employ four Mamba-Transformer blocks. 
Each block uses a hidden size of 768 and 8 attention heads. 
The Mixture-of-Experts (MoE) module includes 4 experts per expert group.
Each Transformer has three Mamba blocks. 
The total number of parameters is 177 M. 
The flow-matching training uses 1,000 timesteps, while inference employs 25 timesteps with the Euler ODE solver. During training, we use 8 NVIDIA RTX-4090 GPUs with a batch size of 12K frames per GPU for 100K steps. The Adam optimizer is applied with a learning rate of $5 \times 10^{-5}$, $\beta_1 = 0.9$, $\beta_2 = 0.999$, and 10K warm-up steps.

\subsection{Training Procedure}
\label{app: train}

As detailed in Section \ref{sec: mpe}, the final loss for the Multimodal Pose Encoder includes:
1) $\mathcal{L}_{contras}$: the contrastive objective for three modalities.
2) $\mathcal{L}_{dur}$: the mean squared error (MSE) duration loss between the predicted and ground truth phoneme-level durations on a logarithmic scale.

In Appendix \ref{app: auto}, the final loss terms for the spatial audio encoder and accomp decoder are as follows:
1) $\mathcal{L}_{rec}$: the L2 reconstruction loss of mel-spectrograms.
2) $\mathcal{L}_{adv}$: the LSGAN-styled adversarial loss for the GAN discriminator.

As for Immersive Drama Transformer, the final loss terms during training include:
1) $\mathcal{L}_{flow}$: the flow matching loss, as described in Section \ref{sec: idt}.
2) $\mathcal{L}_{pitch}$: the MSE pitch loss between the predicted and ground truth f0 in the log scale.
3) $\mathcal{L}_{balance}$: the load-balancing loss for each expert group in Drama-MOE, as discussed in Section \ref{app: moe}.

\subsection{Inference Procedure}
\label{app: infer}

For enhanced generation quality and contextual prosody consistency, we utilize the prompt audio $a$ and the last predicted audio from the same speaker, $y_{pr-last}$, using the Classifier-Free Guidance (CFG) strategy.  
During training, we randomly select prompt audio from the same speaker within the same script to improve generalization, with a 0.4 probability of selecting other audio from the same speaker and a 0.2 probability of dropping the prompt audio entirely.
During inference, we modify the output vector field as in Equation \ref{equ: cfg}.
When $\alpha = 1$, $v_{cfg}$ relies solely on the input prompt audio. 
Furthermore, if $\alpha = 1$ and $\gamma = 1$, $v_{cfg}$ is equivalent to the original $v_t(x, t|a,C;\theta)$. 
Setting $\gamma = 3$ and $\alpha = 0.4$, we improve generation quality and incorporate previously generated audio to enhance the prosody consistency of the same speaker within a single drama act. 
This ensures coherence while preserving the timbre, accent, and articulation of the original prompt audio.
Moreover, as prosody can be learned from previous prompt audio in the same context, this method further enhances prosodic expressiveness and alignment with the drama narrative.

\subsection{Multimodal Pose Encoder}
\label{app: maz}

For inputs from video and geometric pose, we need to segment the length of each speaker transition. In the case of geometric pose, a sudden jump in position indicates a speaker switch, allowing us to determine the total duration of each actor’s line.
For video inputs, we also input the pixel coordinates of the speaker’s lips for the onset of each speech and the corresponding starting frame. 
This information is utilized by Cotracker3 \citep{karaev2024cotracker3} to track the speaker’s mouth movements, facilitating the modeling of 3D position coordinates and quaternion orientation. This process also helps in accurately determining the total duration of each actor’s line.
For duration prediction, we estimate the phoneme durations within each actor’s line based on speaker transitions in the script. 
In addition to encoding semantic and phonetic information as inputs to the duration predictor, we incorporate the length from the video segment and the pose sequence as constraints for the actor’s line duration. 
This ensures a highly precise and consistent duration prediction.

We design three types of contrasts for contrastive learning to explore diverse physical and spatial features. 
Dynamic features include mobility, which differentiates between moving and stationary states, movement speed to capture variations in velocity, and movement direction to account for distinct directional shifts. 
Postural features focus on posture, distinguishing between standing and sitting, and orientation, which captures differences in the relative facing direction of the sound source.
Finally, positional features emphasize varying distances to explore positional relationships and different angles to capture changes in the viewing perspective. Together, these dimensions comprehensively enhance the robustness of our contrastive learning framework.

\subsection{Spatial Audio Encoder and Decoder}
\label{app: auto}

The Spatial Audio Encoder and Decoder are designed based on the Variational Autoencoder (VAE) model \citep{kingma2013auto}. 
During pre-training, we optimize the encoder and decoder using the L2 reconstruction loss. 
To further enhance reconstruction quality, we integrate a GAN discriminator inspired by the architecture of ML-GAN \citep{chen2020hifisinger}. 
Specifically, we employ the LSGAN-style adversarial loss \citep{mao2017least}, $\mathcal{L}_{adv}$, which minimizes the distributional divergence between the predicted mel-spectrograms and the ground truth mel-spectrograms.
Before encoding, we extract the mel-spectrogram using librosa \footnote{\label{foot: librosa}\url{https://github.com/librosa/librosa}}. 
After generating the mel-spectrogram from the decoder's output, HiFi-GAN \citep{kong2020hifi} is used to convert it back into an audio waveform.
To improve the quality of speech generation, we add the reconstructed F0 from our Immersive Drama Transformer during inference, applying the neural source filter (NSF) strategy for enhanced quality.

\subsection{Mamba-Transformer Block}
\label{app: mtb}

\renewcommand{\algorithmicrequire}{\textbf{Input:}}
\renewcommand{\algorithmicensure}{\textbf{Output:}}
\begin{algorithm*}[ht]
\caption{Pseudo-Code of Drama-MOE Routing Strategy}
\label{alg: moe}
\begin{algorithmic}[1]
\REQUIRE Input hidden representation $h$, content embedding $z_c$, prompt audio embedding $z_a$, pose embedding $z_p$, time step $t$
\ENSURE Output with enhanced quality and control $o_{\text{final}}$
\STATE Initialize Gumbel-Softmax temperature $\tau$, sample Gumbel noise $\zeta$
\FOR{each time step $t$}
    \STATE \textbf{Prosidic MOE:}
    \STATE \quad Use Cross-Attention modeling prosody for alignment between $z_a$ and $z_p$:
    \STATE \quad $z_{pro} \gets \text{CrossAttention}(z_p(Q),z_a(K),z_a(V))$
    \STATE \quad Use Gumbel-Softmax for each token in the time channel to select an expert by $z_{pro}$:
    \STATE \quad $g_{\text{prosodic}}(h) \gets \text{GumbelSoftmax}(z_{pro} \cdot W_{\text{prosodic}} + \zeta) / \tau$
    \STATE \quad Compute prosodic MOE output:
    \STATE \quad $o_{\text{prosodic}} \gets \sum_{i} g_{\text{prosodic}, i} \cdot \text{Expert}_{i, \text{prosodic}}(h)$
    \STATE \textbf{Spatial MOE:}
    \STATE \quad Use Gumbel-Softmax for each token in the time channel to select an expert by $z_p$:
    \STATE \quad $g_{\text{spatial}}(o_{\text{prosodic}}) \gets \text{GumbelSoftmax}(z_p \cdot W_{\text{spatial}} + \zeta) / \tau$
    \STATE \quad Compute spatial MOE output:
    \STATE \quad $o_{\text{spatial}} \gets \sum_{i} g_{\text{spatial}, i} \cdot \text{Expert}_{i, \text{spatial}}(o_{\text{prosodic}})$
\ENDFOR
\STATE Return $o_{\text{final}} \gets o_{\text{spatial}}$ as the final routed output
\end{algorithmic}
\end{algorithm*}

The integration of Transformer and Mamba layers provides a flexible framework to balance the often competing objectives of low memory consumption, high computational throughput, and output quality \citep{fei2024dimba}. 
As sequence lengths increase, attention operations progressively dominate computational costs. 
In contrast, Mamba layers are inherently more compute-efficient, and increasing their proportion within the model improves throughput, particularly for longer sequences. 
After experimentation, we determine that an optimal balance is achieved with an attention-to-Mamba ratio of $1:K$, where $K$ is set to 3.

To enhance training stability and prevent numerical instability caused by uncontrolled growth in absolute values, we adopt RMSNorm \citep{zhang2019root}. 
For encoding scene $s$, we encode a video frame or textual scene descriptions into scene embedding $z_s$.
Subsequently, the global embedding $z_g$ is computed by averaging the prompt audio embedding $z_a$ and the scene embedding $z_s$ along the temporal dimension, followed by the addition of the time step embedding $z_t$. This global embedding is processed through a global adaptor, which modulates the latent representation via adaptive layer normalization (AdaLN) \citep{peebles2023scalable} to ensure style consistency. 
The scale and shift parameters are calculated using linear regression:
\begin{equation}
\begin{aligned}
&AdaLN(h, c) = \gamma_c \times LayerNorm(h) + \beta_c,
\end{aligned}
\end{equation}
where $h$ represents the hidden representation. The batch normalization scale factor $\gamma$ is zero-initialized in each block \citep{peebles2023scalable}. 
Additionally, we utilize rotary positional embeddings (RoPE) \citep{su2024roformer} as a form of relative positional encoding, injecting temporal positional information into the model. 
This enhances the model's ability to capture temporal relationships between sequential frames, leading to notable performance gains in the transformer.

Furthermore, a zero-initialized attention mechanism \citep{bachlechner2021rezero} is employed. 
Given the queries $Q_h$, keys $K_h$, and values $V_h$ derived from the hidden states, as well as the scene keys $K_s$ and values $V_s$, the final attention output is computed as:
\begin{equation}
\begin{aligned}
&Attention = \text{softmax} \left( \frac{\tilde{Q_h} \tilde{K_h}^\top}{\sqrt{d}} \right) V_h + \\
&\tanh(\alpha) \text{softmax} \left( \frac{\tilde{Q_h} K_s^\top}{\sqrt{d}} \right) V_s,
\end{aligned}
\end{equation}
where $\tilde{Q_h}$ and $\tilde{K_h}$ incorporate RoPE for queries and keys, $d$ represents the dimensionality of these vectors, and $\alpha$ is a zero-initialized learnable parameter that modulates the cross-attention with the scene embedding.

\subsection{Drama-MOE}
\label{app: moe}

Following previous works \citep{zhang2025versatile}, our routing mechanism employs the dense-to-sparse Gumbel-Softmax technique \citep{nie2021evomoe} to achieve adaptive and efficient expert selection. 
This method utilizes the Gumbel-Softmax trick, which reparameterizes categorical variables to make sampling differentiable, enabling dynamic routing.
For hidden state $h$, the routing score assigned to expert $i$, denoted as $g(h)_i$, is calculated as:
\begin{equation}
\begin{aligned}
&g(h)_i = \frac{\exp((h \cdot W_g + \zeta_i) / \tau)}{\sum_{j=1}^N \exp((h \cdot W_g + \zeta_j) / \tau)},
\end{aligned}
\end{equation}
where $W_g$ is the trainable gating weight, $\zeta$ represents noise sampled from a Gumbel(0, 1) distribution \citep{jang2016categorical}, and $\tau$ denotes the softmax temperature.
At the beginning of training, $\tau$ is set to a high value, promoting denser routing where multiple experts may contribute to processing the same input. As training advances, $\tau$ is gradually lowered, resulting in more selective routing with fewer experts involved. When $\tau$ approaches zero, the output distribution becomes nearly one-hot, effectively assigning each token to the most relevant expert.
Following the approach of \citep{nie2021evomoe}, we gradually decrease $\tau$ from 2.0 to 0.3 during training to transition from dense to sparse routing. During inference, a deterministic routing mode is applied, ensuring that only one expert is chosen for each token. The complete Drama-MOE algorithm is outlined in Algorithm \ref{alg: moe}.

To prevent overloading any single expert and to ensure balanced utilization, we integrate a load-balancing loss for each expert group, as described in Section \ref{sec: idt}, following the approach in \citep{fedus2022switch}. 
For the regularization strength hyperparameter, we set it to 0.1 in our implementation.
The load-balancing mechanism promotes a more uniform allocation of tokens across experts, thereby improving training efficiency by mitigating issues such as expert underutilization or excessive workload. Consequently, our routing strategy not only facilitates dynamic and adaptive expert selection but also ensures an even distribution of computational resources. 
This leads to reduced training time and enhanced performance for the Drama-MOE.

\section{Evaluation Metrics}
\label{app: eva}

\begin{figure*}[ht]
\centering
\includegraphics[width=1\textwidth]{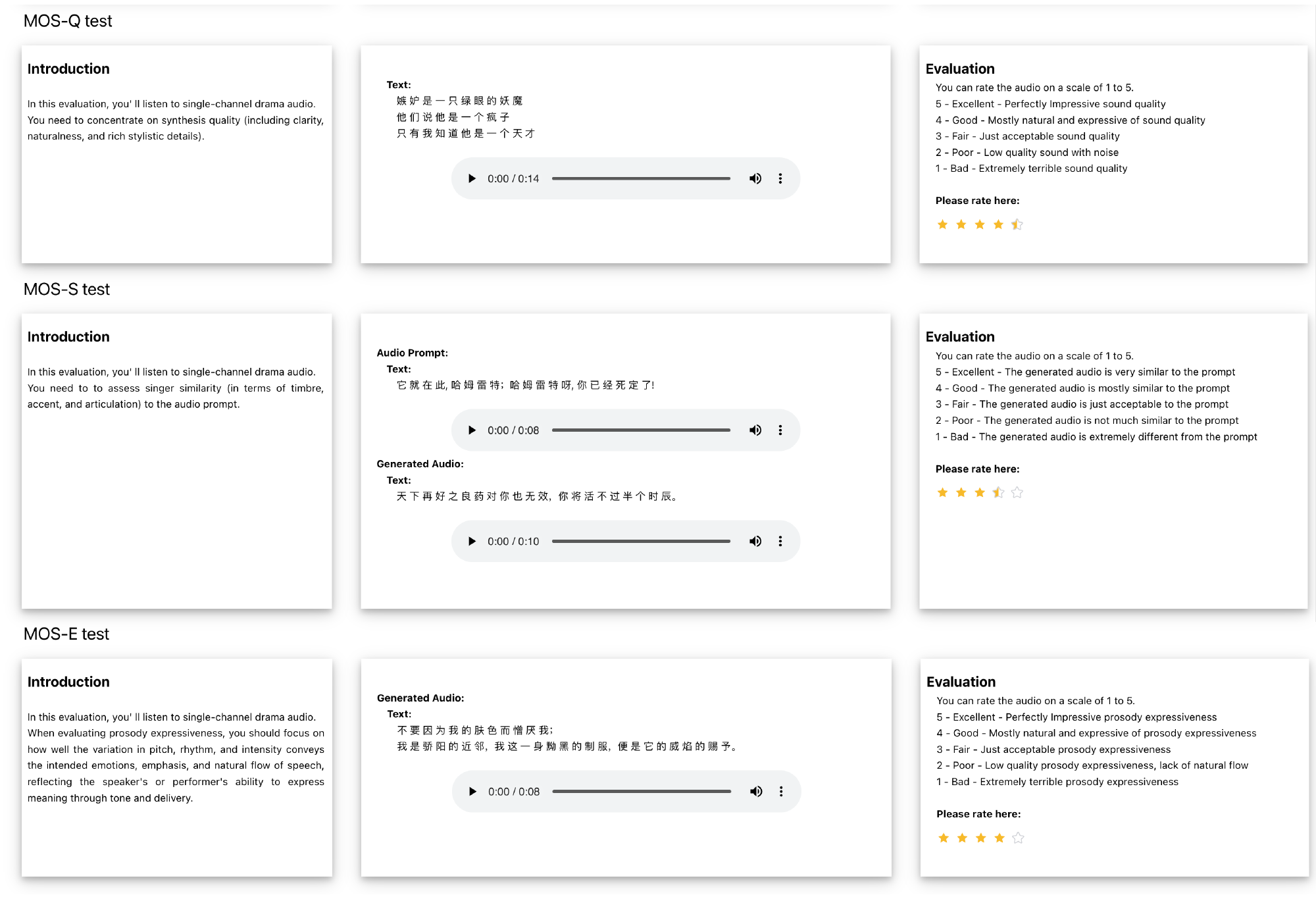}
\caption{Screenshot of monaural speech evaluation.}
\Description{Screenshot of monaural speech evaluation.}
\label{fig: mos1}
\end{figure*}

\begin{figure*}[ht]
\centering
\includegraphics[width=1\textwidth]{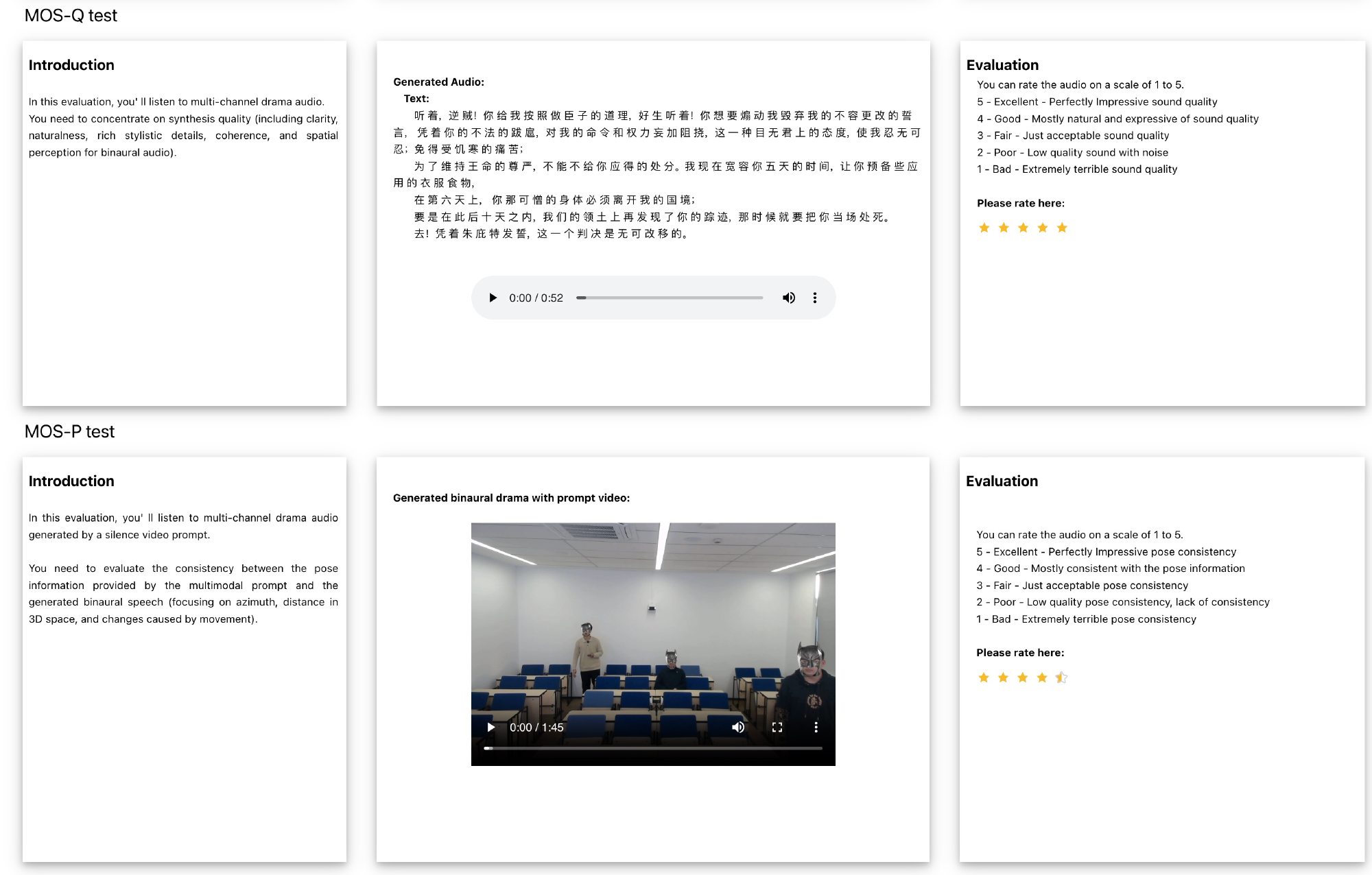}
\caption{Screenshot of binaural speech evaluation.}
\Description{Screenshot of binaural speech evaluation.}
\label{fig: mos2}
\end{figure*}

\subsection{Objective Evaluation}

We evaluate speech intelligibility using the Character Error Rate (CER). The CER is calculated by comparing the transcribed text from speech to the original target text by Paraformer-zh \citep{gao2022paraformer}.

To objectively evaluate speaker similarity, we employ Cosine Similarity (SIM). 
SIM measures the resemblance in speaker identity between the synthesized speech and the GT speech by computing the average cosine similarity between the embeddings extracted from the synthesized and GT speech, thus serving as an objective measure of speaker similarity.
We use WavLM \citep{chen2022wavlm} model fine-tuned for speaker verification \footnote{https://huggingface.co/pyannote/speaker-diarization} to extract speaker embeddings.

We employ F0 Frame Error (FFE) to evaluate the synthesis prosody of the test set objectively. FFE combines metrics for voicing decision error and F0 error, capturing essential synthesis quality information.

For the objective evaluation of IPD and ILD, we first convert the time-domain signal $x(n)$ into the frequency-domain signal $X(t,f)$ using the short-time Fourier transform (STFT):
\begin{equation}
\begin{aligned}
&X_i(t,f) = \sum_{n=0}^{N-1} x_i(n) \cdot w(t-n) \cdot e^{-j2\pi fn}, i \in \{1, 2\},
\end{aligned}
\end{equation}
where \( w(t-n) \) is a window function, \( N \) is the window length, and $i$ indicates the channel of the binaural audio.
Next, we calculate the mel-spectrogram, IPD, and ILD based on the frequency-domain signals $X_i(t,f)$. 
The mel-spectrogram for each channel is calculated as:
\begin{equation}\label{mel-eq}
\begin{aligned}
&S_i(t, m) = \log\left(|X_i(t, f)|^2 \times \text{melW} \right),
\end{aligned}
\end{equation}
where melW is an $M$-bin mel filter bank.
The IPD is derived from the phase spectrograms of the left and right channels:
\begin{equation}\label{ipd-eq}
\begin{aligned}
&IPD(t, f)=\angle \frac{X_2(t, f)}{X_1(t, f)}.
\end{aligned}
\end{equation}
ILD is extracted from the loudness spectrum of the left and right channels:
\begin{equation}\label{ild-eq}
\begin{aligned}
& ILD(t, f)=20\log_{10}\left( \dfrac{|X_2(t,f)| + \varepsilon}{|X_1(t,f)| + \varepsilon} \right), \varepsilon = 1e^{-10}.
\end{aligned}
\end{equation}
We calculate Mean Absolute Error (MAE) metrics based on the IPD and ILD extracted from the ground truth (GT) and the predicted speech. 
Since the IPD here is in radians and the ILD uses log10, the resulting values are quite small, especially after averaging the MAE over the time dimension. 
So, we multiply by 100 to make the results more intuitive.

Additionally, we analyze angular and distance metrics using SPATIAL-AST \citep{zheng2024bat}. 
SPATIAL-AST encodes an angle and a distance embedding for binaural audio. 
We compute and average the cosine similarity for each 1-second segment based on the GT and predicted audio.

\subsection{Subjective Evaluation}

We conduct Mean Opinion Score (MOS) as a subjective evaluation metric. For each task, we randomly select 40 pairs of sentences from our test set for subjective evaluation. Each pair consists of an audio prompt that provides timbre, and a synthesized speech sample, both of which are evaluated by at least 15 professional listeners.

In the context of MOS-Q evaluations, these listeners are instructed to concentrate on synthesis quality (including clarity, naturalness, rich stylistic details, coherence, and spatial perception for binaural audio). Conversely, during MOS-S evaluations, the listeners are directed to assess speaker similarity (in terms of timbre, accent, and articulation) to the audio prompt. For MOS-E, the listeners are informed to evaluate prosodic expressiveness. For MOS-P, the listeners are instructed to evaluate the consistency between the pose information provided by the multimodal prompt and the generated binaural speech (focusing on angle, distance in 3D space, and changes caused by movement).

In MOS evaluations, listeners are requested to grade various speech samples on a Likert scale ranging from 1 to 5. 
Notably, all participants are fairly compensated for their time and effort. We compensated participants at a rate of \$12 per hour, with a total expenditure of approximately \$300 for participant compensation. Participants are informed that the results will be used for scientific research. The instructions for testers in monaural and binaural evaluation are shown in Figure \ref{fig: mos1} and Figure \ref{fig: mos2}.

For CMOS, listeners are asked to compare pairs of audio generated by systems A and B and indicate their preference between the two. They are then asked to choose one of the following scores: 0 indicating no difference, 1 indicating a slight difference, 2 indicating a significant difference, and 3 indicating a very large difference.

\begin{figure*}[ht]
\centering
\includegraphics[width=1\textwidth]{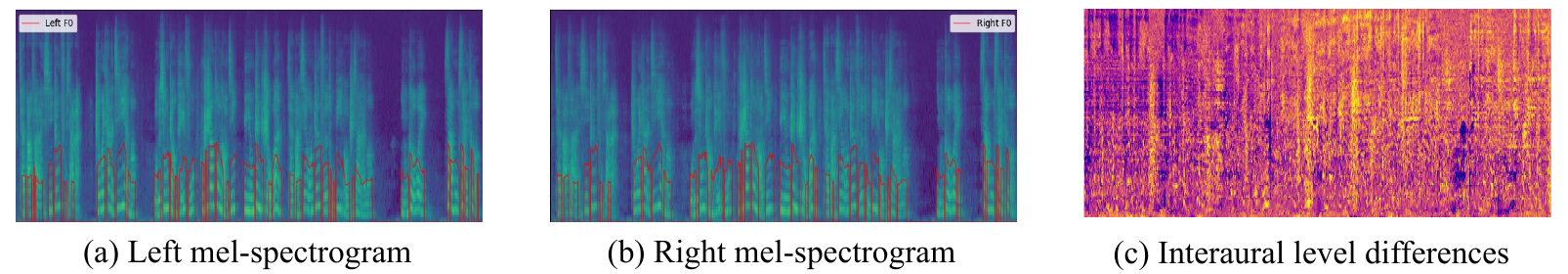}
\caption{Visualization results.}
\Description{Visualization results.}
\label{fig: exp}
\end{figure*}

\section{Extended Experiments}
\label{app: exp}

\subsection{Baseline Models}
\label{app: base}

\textbf{UniAudio} \citep{yang2023uniaudio} uses LLMs to generate multiple audio types by tokenizing target audio and conditions, concatenating them into a single sequence, and performing next-token prediction. A multi-scale Transformer is introduced to handle long sequences caused by neural codec-based VQ.
We employ their official code \footnote{https://github.com/yangdongchao/UniAudio}.

\textbf{StyleTTS 2} \citep{li2024styletts} combines style diffusion and adversarial training with large speech language models (SLMs) for high-quality text-to-speech synthesis. It models style as a latent random variable using diffusion models.
We employ their official code \footnote{https://github.com/yl4579/StyleTTS2}.

\textbf{CosyVoice} \citep{du2024cosyvoice} represents speech with supervised semantic tokens derived from a multilingual speech recognition model, using vector quantization in the encoder. It employs an LLM for text-to-token generation and a conditional flow matching model for token-to-speech synthesis.
We employ their official code \footnote{https://github.com/FunAudioLLM/CosyVoice}

FireRedTTS \citep{guo2024fireredtts} is a language-model-based TTS system that encodes speech into discrete semantic tokens via a semantic-aware speech tokenizer. 
A language model then generates these tokens from text and audio prompts, followed by a two-stage waveform generator for high-fidelity synthesis.
We employ their official code \footnote{https://github.com/FireRedTeam/FireRedTTS}.

\textbf{F5-TTS} \citep{chen2024f5}, a non-autoregressive text-to-speech system based on flow matching with Diffusion Transformer (DiT). Text input is padded with filler tokens to match the length of input speech, followed by denoising for speech generation.
We employ their official code \footnote{https://github.com/SWivid/F5-TTS}.

\subsection{Visualization Results}

Figure \ref{fig: exp} presents a visualization of our binaural speech synthesis results.
In Figures (a) and (b), we illustrate the mel spectrogram and F0 contour of a single speech sample for the left and right audio channels. 
ISDrama effectively captures fine-grained mel details and generates expressive F0 variations, demonstrating high-quality synthesis and rich prosody.
Figure (c) visualizes the Interaural Level Difference (ILD). 
The darker coloration on the left side indicates that the sound source initially originates from the back-left position, while the lighter coloration on the right side corresponds to the source moving toward the back-right. 
This clearly shows that ISDrama successfully models positional information, accurately reflecting spatial dynamics.

\subsection{Context-Consistent CFG}

We conduct experiments to verify the parameter settings of Equation \ref{equ: cfg}, as shown in Table \ref{tab: cfg}.
For evaluation, we implement CMOS assessments.
When $\alpha = 1$, $v_{cfg}$ relies solely on the input prompt audio. Moreover, if $\alpha = 1$ and $\gamma = 1$, $v_{cfg}$ becomes equivalent to the original formulation $v_t(x, t|a,C;\theta)$.

\begin{table}[ht]
\centering
\caption{Ablation study for context-consistent CFG.}
\label{tab: cfg}
\begin{tabular}{lcccc}
\toprule
$ \bf \alpha$ & \bf $\gamma$ & \bf CMOS-Q & \bf CMOS-S & \bf CMOS-E \\
\midrule
1.00 & 1.00 & -0.62 & -0.09 & -0.78 \\
0.40 & 1.00 & -0.31 & -0.16 & -0.43 \\
0.40 & 1.50 & -0.36 & -0.22 & -0.51 \\ 
1.00 & 3.00 & -0.28 & -0.02 & -0.55 \\ 
0.40 & 3.00 & \textbf{0.00} & \textbf{0.00} & \textbf{0.00} \\ 
0.00 & 3.00 & -0.30 & -0.24 & -0.22 \\
0.40 & 5.00 & -0.49 & 0.04 & -0.60 \\ 
\bottomrule
\end{tabular}
\end{table}        

When $\gamma$ increases from 1 to 1.5, the generated speech exhibits inconsistencies between the pronunciation and prosody of the prompt audio, resulting in lower CMOS-S and suboptimal CMOS-E.
When $\gamma$ ranges from 1.5 to 3, the generated speech aligns with the pronunciation and accent of the prompt audio. 
Within this range, if $\alpha$ is appropriately set, the model effectively utilizes the previous prompt audio to synthesize semantically aligned dramatic prosody.
When $\gamma > 5$, the prosody adheres to the style of the prompt audio but reduces the ability to adapt to semantic learning.
On the other hand, when $\alpha$ approaches 0, the CMOS-S decreases as the model over-relies on contextual cues, neglecting the fine-grained details of the accent and pronunciation in the prompt audio.
Conversely, when $\alpha$ approaches 1, the CMOS-E decreases because the model becomes overly dependent on the prompt audio, making it challenging to model semantically aligned dramatic prosody.

By setting $\gamma = 3$ and $\alpha = 0.4$, we achieve improved generation quality and incorporate previously generated audio to enhance the prosody consistency of the same speaker within a single dramatic act.

\end{document}